\documentclass{article}

\usepackage[english]{babel}

\usepackage[letterpaper,top=2cm,bottom=2cm,left=3cm,right=3cm,marginparwidth=1.75cm]{geometry}

\usepackage{amsmath}
\usepackage[utf8]{inputenc}
\usepackage{graphicx}
\usepackage[colorlinks=true, allcolors=blue]{hyperref}
\usepackage{xcolor}
\usepackage{color, colortbl}
\definecolor{Gray}{gray}{0.9}
\usepackage{authblk}
\usepackage{eurosym}
\usepackage{float}
\floatstyle{plaintop}    
\usepackage{rotating}
\usepackage{caption}
\usepackage{multirow}
\usepackage[doublespacing]{setspace}
\usepackage{subfiles}
\usepackage{makecell}
\usepackage{natbib}
\setcitestyle{open={[},authoryear,close={]}}
\usepackage{caption}
\usepackage{changepage}
\usepackage{lineno}
\usepackage{forest}
\usepackage{tikz}
\usetikzlibrary{arrows.meta, quotes}
\usepackage{wasysym}

\usepackage{adjustbox}
\usepackage{graphicx}
\usepackage{booktabs,caption}
\usepackage[flushleft]{threeparttable}
\restylefloat{table}
\usepackage{booktabs}
\usepackage{dcolumn}
\newcolumntype{d}[1]{D..{#1}}

\title{Explaining Apparently Inaccurate Self-assessments of Relative Performance: A Replication and Adaptation of ”Overconfident: Do you put your money on it?” by Hoelzl \& Rustichini (2005)}

\author[1]{Marius Protte}
\affil[1]{{\small Paderborn University, Heinz-Nixdorf-Institute, Fürstenallee 11, 33102 Paderborn}}
\date{2025}

\begin{document}
\maketitle

\onehalfspacing
\begin{abstract}

This study replicates and adapts the experimental framework of Hoelzl and Rustichini (2005), which examined overplacement, i.e., overconfidence in relative self-assessments, by analyzing individuals’ voting preferences between a performance-based and a lottery-based bonus payment mechanism. The original study found underplacement -- the majority of their sample apparently expected to perform worse than others -- in difficult tasks with monetary incentives, contradicting the widely held assumption of a general human tendency toward overconfidence. This paper challenges the comparability of the two payment schemes, arguing that differences in outcome structures and non-monetary motives may have influenced participants' choices beyond misconfidence. 
In an online replication of their experiment, a fixed-outcome distribution lottery mechanism with interdependent success probabilities and no variance in the number of winners -- designed to better align with the performance-based payment scheme -- is compared against the probabilistic-outcome distribution lottery used in the original study, which features an independent success probability and a variable number of winners. The results align more closely with traditional overplacement patterns than underplacement, as nearly three-fourths of participants prefer the performance-based option regardless of lottery design. Key predictors of voting behavior include expected performance, group performance estimations, and sample question outcomes, while factors such as social comparison tendencies and risk attitudes play no significant role. 
Self-reported voting rationales highlight the influence of normative beliefs, control preferences, and feedback signals beyond confidence. These results contribute to methodological discussions in overconfidence research by reassessing choice-based overconfidence measures and exploring alternative explanations for observed misplacement effects.


\vspace{1cm}

\end{abstract}

\doublespacing







\doublespacing

\section{Introduction} \label{introduction}

Overconfidence was famously labeled the "[perhaps] most robust finding in the psychology of judgment" by \citet[p. 389]{debondt1995}, and is frequently cited as a key driver of inefficiencies in human decision-making and resulting economic outcomes. The related literature attributes a wide range of financial, managerial, and societal phenomena -- at least in part -- to humans' overconfidence regarding their skills, capabilities, judgments or prospects.
In financial markets, overconfidence has been linked to excessive trading volume and heightened market volatility \citep{odean1998,daniel2001,statman2006}, with diminished trading performance \citep{biais2005}, overreactions to private information and underreactions to public information, increasingly aggressive trading after past gains, underestimation of risk, and investment in riskier securities \citep{chuang2006}. In corporate finance and management, CEOs identified as overconfident are more likely overinvest in internal funds \citep{malmendier2005}, pursue mergers and acquisitions at an elevated rate \citep{malmendier2008}, and engage in self-serving attributions of company performance \citep{libby2012}. Overconfidence has also been linked to the emergence of speculative price bubbles \citep{scheinkman2003}, premature or excessive market entry in highly competitive industries \citep{camererlovallo1999,cain2015,voros2024}, and a reluctance to make necessary strategic adjustments \citep{kraft2022,gervais1998}, ultimately increasing the risk of entrepreneurial failure \citep{bernardo2001,simon2000}\footnote{At the same time, (over)confidence is considered a major driver of success in investing \citep{wang2001}, attaining positions of influence \citep{anderson2012,radzevick2011}, becoming an entrepreneur \citep{townsend2010} and fostering innovation \citep{baek2019}. These associations may be explained by overconfident individuals appearing more competent than those who are objectively competent \citep{anderson2012}, them having propensity to exert greater effort \citep{landier2003}, or greater willingness to take the necessary risks to realize desired outcome in competitive contexts with overconfidence counteraction inefficient risk-aversion \citep{kahneman1993}, enabling individuals to pursue opportunities they might otherwise forgo.}.
Beyond financial and managerial contexts, overconfidence has been implicated in various labor market outcomes \citep{santospinto2020} and broader social and political dynamics. It has been associated with self-deception in decision-making \citep{benabou2002}, under-insurance and inadequate risk protection \citep{sandroni2007}, failed marriages \citep{mahar2003}, and the outbreak of wars \citep{johnson2011}.


 

While early research -- explicitly or implicitly -- treated overconfidence as one uniform psychological concept \citep{moore2008}, a more nuanced understanding has since been established. Typically, literature cites three specific phenomena that are amalgamated under the term "overconfidence": overestimation, overplacement, and overprecision. Regarding one's actual ability, performance, control, or chance of success, \textit{Overestimation} describes a self-assessment that exceeds the level that is objectively justified; \textit{Overplacement} means exaggerating one's placing relative to a reference population, implying a belief to be better than others; \textit{Overprecision} refers to excessive certainty regarding the accuracy of one’s beliefs \citep{moore2008}\footnote{\citet{moore2008} criticize researchers' habits of treating overestimation, overplacement, and overprecision as "interchangeable manifestations of self-enhancement" (p. 503) despite them being "conceptually and empirically distinct" (p. 515) types of overconfidence.}. 
The type of individuals' inaccuracy in self-assessments is typically observed to be confounded with the difficulty of the task on which the assessment is based. Multiple studies show that overplacement occurs on easy tasks, while underplacement occurs on difficult tasks, \citep{healy2007,moore2007cain,moore2007small,moore2008,grieco2009}. The interrelation between overestimation and underestimation runs counter-directional, with overestimation occurring on difficult tasks and underestimation on easy ones \citep{healy2007,grieco2009}. Also, low performance in absolute terms is commonly found to facilitate underplacment on difficult tasks, while high performance are found to be associated with overplacement \citep{kruger1999,brown2016}. 
Meanwhile, estimation and placement are commonly observed to be negatively correlated with respect to task difficulty. Overplacement typically occurs on easy tasks, while overestimation takes place on difficult tasks \citep{healy2007,moore2008,grieco2009}. Overprecision is positively correlated to both other overconfidence types in both directions, i.e. the more precise a self-assessment the less over- or underestimation and over- or underplacement occurs \citep{moore2008}. 
Overplacement\footnote{Overplacement \citep{larrick2007} is closely related to the "better-than-average effect" \citep{alicke2005}. The two terms are often used interchangeably in the literature \citep{moore2008} and will be treated as such in this study, for the sake of conciseness. However, hereafter only "overplacement" will be used, as it is the broader term for inaccuracies in relative self-assessments.} -- and underplacement by association -- are the paradigms of interest for the scope of this study.




A central effort for distinguishing under which circumstances overplacement and underplacement occur respectively was made by Erik Hoelzl and Aldo Rustichini in 2005. They introduce a novel experimental approach to studying overplacement under economic incentives and varying task difficulty in their well-cited original research paper “Overconfident: Do you put your money on it?”. 
By letting participants vote between a performance- based (the best performing half in a knowledge test wins) and a lottery-based (half of potential outcomes in an individual die roll wins) bonus payment mechanism -- both offering an 50\% win probability in expectation –- they use this choice as a behavioral measure for inaccurate self-assessments, with significant miscalibration in either direction being interpreted as overconfidence or underconfidence respectively. This design goes beyond verbal self-assessments of individuals’ own skills and abilities, which were widely used in prior experiments despite being prone to subjective interpretation \citep[e.g., driving skills in][]{svenson1981}.
Notably, H\&R observe a tendency toward underplacement rather than overplacement when a task is difficult and real (as opposed to hypothetical) bonus payments are provided. This finding contrasts with the prevailing assumption of a general and robust human tendency toward overconfidence, as expressed by \citet{debondt1995}, among others. 
Since H\&R, numerous experimental studies on overconfidence have adopted and refined similar performance-based versus conditional or random payment approaches to studying overplacement \citep[see e.g.,][]{blavatskyy2009,urbig2009,grieco2009,park2010,ericson2011,benoit2014,owens2014,koellinger2015,hollard2016, benoit2022}.

Meanwhile, legitimate skepticism has been raised about whether misconfidence can actually be inferred from over- or underplacement data, or from choice behavior in general \citep{benoit2011}. \citet{benoit2014} highlight the mixed evidence on overplacement and emphasize its already limited capacity to support generalizable conclusions within the traditional paradigm, irrespective of the criticism by \citet{benoit2011}.
The debate revolves around three main aspects. 
First, overplacement and underplacement observed in ranking experiments that rely on a single reference point -- typically the median -- and employ a dichotomous categorization of success to infer miscalibration in relative abilities can often be rationalized \citep{benoit2011}. This applies to H\&R's findings as well. Consequently, it is argued that such experiments reveal only "apparent", rather than "true", overconfidence or underconfidence. This distinction is particularly important, as true overconfidence may lead to the aforementioned negative real-world consequences, whereas apparent overconfidence is unlikely to have such effects  \citep{benoit2011,benoit2014}.
Second, overconfidence may be better understood as a statistical bias stemming from information asymmetries or limited information availability rather than as a behavioral or psychological bias \citep{healy2007,grieco2009,benoit2011}. 
Third, alternative decision motives -- such as a preference for control -- may cause overplacement data in choice experiments where individuals bet on their own performance versus random devices, which is misinterpreted as overconfidence \citep{owens2014,benoit2022}. 
Nevertheless, some studies \citep[e.g.,][]{merkle2011,benoit2014} observe behavioral patterns that appear overconfident despite fulfilling \citet{benoit2011}'s criteria for identifying 'true' overconfidence. Thus, prior overconfidence research should not be automatically dismissed or deemed invalid \citep{merkle2011}.
As this debate remains ongoing and extends beyond the scope of this study, key aspects of the discussion will be further outlined in Appendix \ref{validity}. This study primarily aims to evaluate the internal validity of H\&R's experimental results concerning the underlying motives behind subjects’ voting behavior and self-assessment accuracy, rather than to assess their interpretation within the broader overconfidence paradigm. However, this replication will apply due caution in both the use of terminology and the interpretation of results, acknowledging the limitations outlined by the aforementioned research.


H\&R infer underconfidence from a majority of participants preferring the lottery in their \textit{Difficult x Money} treatment condition. Their identification strategy is based on the assumptions of individuals having an incentive to truthfully reveal their self-assessment relative to the group by voting the option they believe maximizes their chance of receiving a bonus: selecting the performance-based payment scheme if they expect to rank in the upper half and opting for the luck-based die roll if they anticipate performing below the group median. 
Despite H\&R’s surprising results and their frequent citations, little attention has been paid to the characteristics of the lottery itself, even though the issue of alternative explanations for voting behavior has been discussed in the literature. This study argues that the two payment scheme alternatives are not fully comparable, as they differ in their degree of outcome dependency and allow for inconsistent numbers of participants receiving a bonus.
Under monetary incentives and difficult tasks, these structural differences may alter voting behavior through mechanisms such as aversion to social comparison, low self-efficacy, prosociality or other non-monetary motives, potentially shifting votes in favor the lottery option.
Therefore, this replication examines the robustness of H\&R's results to a modified lottery mechanism that better aligns with the characteristics of the performance tests. Additional controls explore whether miscalibrated voting distributions, in fact, stem from inaccurate self-assessments or other underlying factors.

The importance of replications in experimental economic research -- much like in social sciences in general -- has been emphasized since the field's inception \citep[see e.g.,][]{smith1970,rosenthal1990,lamal1990,charness2010}. By replicating H\&R’s experiment and examining whether the characteristics of the lottery device themselves may influence behavior, this study contributes to the ongoing methodological discussions on the feasibility of using performance-based versus random-device-based choices in experimental overconfidence research. It also contributes to developing a more comprehensive understanding of misplacement in individual self-assessments and behavioral factors besides misconfidence that may be mistakenly interpreted as such.
Beyond methodological implications, the findings may also have practical relevance for employees’ compensation plan choices and organizations' compensation plan design \citep{brown2016}. 

The replication results show voting distributions that align more closely with the traditional understanding of overplacement than with the underplacement observed in H\&R. The alternative lottery mechanism does not appear to influence voting behavior. Additionally, four key factors driving individuals' voting decisions are identified: confidence level, signals obtained through sample questions, normative beliefs, and -- exclusively among test voters -- a preference for control.

The remainder of this paper is structured as follows: Section \ref{experiment} provides a brief summary of the original study by H\&R, outlining its research question, contribution, experimental design, and key findings. Section \ref{hypotheses} then discusses the limitations of H\&R’s study and derives the hypotheses for the present replication study. Subsequently, the experimental design and experimental results will be presented in detail, following the structure of H\&R’s paper with selective modifications. Section \ref{experiment_new} introduces task selection, sampling, experimental procedure, treatment variations, and the incentive scheme. Section \ref{results} presents descriptive and non-parametric analyses of the experimental results, examining subjects’ predictions, voting behavior, and identifying predictors and correlates of their voting decisions. Additionally, an exploratory examination of subjects’ self-reported voting rationales will be conducted. Finally, Section \ref{discussion} summarizes the study’s key findings and concludes with an in-depth discussion of its limitations and implications.

\section{Original Experiment by Hoelzl \& Rustichini (2005)} \label{experiment}


Instead of relying on verbal self-reports to measure subjective beliefs, as was commonly practiced at the time, Eric Hoelzl and Aldo Rustichini (hereinafter "H\&R") present a novel approach to studying overconfidence in relative comparisons, i.e. overplacement. Their original experiment introduces a voting decision with payoff implications, serving as a more objective behavioral measure of overplacement relative to a reference group. The experimental design required subjects to indicate their preference between two distinct bonus payment allocation mechanisms -- a performance-based one and a lottery-based one -- by casting a vote. The mechanism that received the majority of votes was then applied to all participants in the respective group.

H\&R employ a 2x2 experimental design, differentiating between real monetary and hypothetical incentives, as well as between easy and difficult tasks. While potential economic implications of overconfidence had been extensively discussed before, H\&R were among the first experiments on overconfidence to incorporate monetary incentives. In the monetarily incentivized treatments, 50\% of participants -- in expectation -- would receive a bonus payment. The allocation of this bonus was determined either by the performance in a vocabulary test (with the top  50\% receiving the bonus) or by an individual die roll, where half of the possible outcomes (rolling a 4, 5, or 6) awarded the bonus\footnote{In the treatments with hypothetical incentives, subjects were asked to imagine the respective payoff structure.}. Since real-life economic decisions are typically constituted by monetary stakes, the monetary incentive conditions should be considered the more relevant of the four treatments. As a second treatment manipulation, besides the nature of the incentive (real vs. hypothetical), task difficulty varied between groups. Test items were explicitly labeled as either easy or difficult and were designed accordingly.

An individual's voting choice served as a proxy for their confidence level relative to the group. Consequently, overplacement was inferred from a majority of participants preferring the performance test-based payoff, whereas a majority favoring the lottery-based mechanism was considered underplacement. H\&R argue that voting for the performance-based mechanism is a weakly dominant strategy for anyone who believes they are more skilled than the median participant, assuming participants aim to maximize their personal payoffs. Therefore, the voting mechanism should ensure a truthful revelation of subjects' skill self-evaluation\footnote{For a more extensive discussion of the game's equilibria see 1.1.3 (p. 310) in H\&R as well as their Technical Appendix under A.3.}.

The experimental procedure was conducted using a seven-page questionnaire that was distributed and collected sequentially. Subjects were recruited from the university campus and participated in groups of at least seven. Before casting their vote on their preferred payoff mechanism, subjects were provided with two sample items and their respective solutions. Participants then completed both the knowledge test \textit{LEWITE} ("Lexikon Wissenstest") \citep{wagner1999}\footnote{This test featured a mechanism for comparing self-assessment with an immediate knowledge check. First, participants indicated which of the twenty listed words they knew and could explain. Second, they were presented with explanation sentences for all twenty words and were required to fill in two blanks for each definition by selecting from seven to nine answer options. The structure of the test provided an additional measure of inaccurate self-assessment beyond the vote and made success through guessing highly unlikely.} and the lottery task, without knowing which payoff scheme would ultimately be implemented until the experiment’s conclusion. This design choice aimed to prevent biases in voting behavior arising from procedural preferences, such as effort avoidance\footnote{For a general discussion of the effect of equal payments on effort see \citet{abeler2010gift}.}. Participants indicated their expectations for the test beforehand and reflected on it afterward through standardized Likert-scale questions. Additionally, they were asked to predict their own test performance in advance and estimate both their own and the group’s scores after completing the test.

The experimental results notably do not support the presumption of a robust general human tendency for overconfidence in self-assessments, as voiced by \citet{debondt1995}. While H\&R observe the majority of subjects favoring the performance-based bonus allocation scheme in three out of their four groups, the \textit{Difficult x Money} treatment stands out from the others. Under real monetary incentives with hard test items, the majority vote shifts toward the luck-based payoff scheme (die roll), presumably because subjects perceive their chances of obtaining a bonus to be higher in a fair 50/50 lottery than in a difficult knowledge test. H\&R infer overplacement tendencies for both easy and difficult tests under hypothetical incentives. Only when real monetary incentives are provided do individuals display a tendency for underplacement in relative self-assessments. Logit regression analysis of voting for the test on the expectation of performing better than the group average reveals an overall positive coefficient that becomes insignificant for the difficult test with money at stake. It is therefore concluded that a combination of monetary incentives and a difficult task discourages individuals from choosing a performance-based payoff scheme, even when they anticipate outperforming their competition.

\section{Limitations to H\&R and Derivation of Hypotheses} \label{hypotheses}

More recent literature on performance- vs. random device-based payoff scheme choices has hinted that the voting patterns observed in H\&R may not necessarily indicate underconfidence but rather reflect aversion to risk or ambiguity associated with the test \citep{owens2014,blavatskyy2009,grieco2009}.
H\&R themselves acknowledge that under economic incentives with a difficult test their measure may be skewed toward the lottery. While the lottery features a clearly defined win probability, the capabilities of one's competitors in the performance test -- and the resulting probability of winning -- are ambiguous. However, individuals' willingness to sacrifice potential earnings or pay an insurance premium to avoid ambiguity has been pointed out in the literature \citep[see e.g.,][]{ellsberg1961,hogarth1989,fox1995,moore2003}, particularly when stakes are high \citep{blavatskyy2023}. 
Meanwhile, ambiguity attitudes are typically confounded with risk preferences \citep{gneezy2015}, which H\&R abstract from in their equilibrium derivation of subjects' voting strategy. Consequently, the extent to which underconfidence drives the shift toward the lottery may be overestimated, with some of the observed voting behavior instead being explained by aversion against ambiguity and risk.


Beyond these limitations, which H\&R mention in passing, little attention has been paid toward the structural differences of the test and lottery mechanisms. This study argues that the two payoff schemes are not fully equivalent, as their statistical properties allow for different distributions of bonus recipients  while simultaneously exhibiting differing degrees of outcome-dependency. 
The performance test determines a fixed number of winners (with no variance: $\sigma^2=0$), which is known to participants ex ante given knowledge about group size, with individual win probabilities being interdependent. In contrast, an individual die roll provides each participant with an independent 50\% chance of winning. Therefore, while the lottery assigns 50\% winners in expectation, it introduces a variance to the actual number of winners ($\sigma^2=n/4$), which, in the most extreme -- though unrealistic -- case, result in every single participant receiving a bonus.

The anticipation of this flexibility in the number of potential bonus recipients under the lottery scheme -- in conjunction with the independence of win probabilities -- may create a discrepancy in how the two mechanisms are perceived, particularly given the prospect of real monetary incentives and a difficult task. While the test condition constitutes a zero-sum environment -- where a fixed number of bonuses are awarded and one participant’s success necessarily precludes another’s -- the lottery does not impose such constraints. 
Thus, voting for the lottery option, which provides each group member with an independent and objective 50\% chance of receiving the bonus, holds the potential to improving social welfare \citep{charness2002} beyond what would be possible under the performance-based scheme.
This reasoning aligns with findings from redistribution research, which suggests that "inequality is tolerated more if it does not come at the expense of others" \citep[][p. 2]{strang2025}.
Although a minority of individuals may be classified as purely selfish payoff-maximizers, most people exhibit at least some degree of altruism or fairness concerns and are often willing to forgo personal monetary gains to either benefit others directly or reduce payoff inequalities \citep{andreoni2002,loewenstein1989,bolton2000,fehr1999,engelmann2004,fehr2006}.
As a result, voting behavior may deviate from subjects' relative self-assessments, with some participants who could reasonably expect to place in the upper half of the group in terms of performance still opting for the lottery. This pattern should be less pronounced under hypothetical payment schemes or easier tasks with lower performance differentiation.
Conversely, inequality based on performance is often considered acceptable \citep{freyer2023inherited,gee2017redistributive}, but primarily when the exerted effort is observable, as this allows people to distinguish between outcomes due to effort versus those due to luck or talent \citep{fischbacher2017nonadditivity}, while people are more likely to attempt to mitigate inequality perceived to stem from luck \citep{mollerstrom2015luck,rey-biel2015}. This concern is arguably more relevant to the lottery mechanism -- somewhat ironically, given that it is based on luck and could also result in fewer than 50\% of participants winning -- as it allows for variance in the number of bonus recipients and may thereby reduce inequality, unlike the test condition.
\citet{krawczyk2012} notes that when bonus payments are conditional on success in a task or lottery, some individuals may prefer a test-based bonus allocation because they consider performance-dependent payment fair, while others may favor a lottery for its ability to reduce outcome inequality. 




Accordingly, votes for the lottery-based payoff mechanism may reflect a broader and more diverse array of motives -- particularly non-monetary ones -- beyond purely reflecting relative self-assessment under payoff-maximizing constraints outlined in H\&R's equilibrium strategy derivation. Their identification strategy relies on the assumption that subjects have an incentive to truthfully reveal their self-assessment relative to the group through their vote, as it goes along with the highest chance of receiving the bonus. 
Consequently, H\&R’s results likely overestimate participants’ actual self-assessment-induced preference for the lottery-based payoff mechanism and, in turn, the extent of underconfidence.

Implementing an alternative lottery mechanism that converts the die roll into a random device with a fixed-outcome distribution and interdependent success probabilities should objectively equalize potential perceived disparities in the numbers of winners between the performance test and the lottery\footnote{The modified lottery mechanism however does not affect the ambiguity associated with the test.}. This adjustment may lead to voting results that more accurately capture individuals' subjective relative self-assessment, aligning more closely with the study’s original intent.

Thus, the following hypotheses are proposed for comparing H\&R’s original experiment with the modified lottery mechanism:\\
\\
\textbf{Hypothesis 1:} Voting behavior differs between a lottery with an independent win probability of 50\% and a lottery with a fixed-outcome distribution win rate of 50\%. 
\\
\\
\textbf{Hypothesis 2:} Specifically, less underplacement occurs with a fixed-outcome distribution lottery compared to the probabilistic lottery used in H\&R.




\section{Current Experiment} \label{experiment_new}



For this replication, conducted in late 2024 via the academic crowd-working platform Prolific\footnote{Prolific was chosen as it allows for the application of specific sampling criteria  \citep{palan2018prolific}, and its participant pool has been shown to be more attentive than those on other platforms such as MTurk \citep{albert2023comparing}. For a more detailed comparisons with other platforms, see \citet{peer2017beyond}.}, H\&R's experiment was translated into an online format using oTree \citep{chen2016}. The study was pre-registered in the American Economic Association's registry for randomized controlled trials and was approved by the German Association for Experimental Economic Research e.V. (GfeW)\footnote{AEA RCT Registry in May 2024: https://doi.org/10.1257/rct.13070-1.0. \par GfeW Registry in February 2024: https://gfew.de/ethik/4dqGpmS5} prior to the experiment.
The replication focuses solely on the \textit{Difficult x Money} group from H\&R, which includes real monetary incentives and a difficult task, as it is the only condition in which behavior significantly differed from the other groups, while also being the most relevant from an economic perspective. The full experimental instructions are available in Appendix \ref{instructions}, while a detailed comparison by aspect between this experiment and H\&R's original version can be found in Appendix \ref{appendix}.

\subsection{Task}

Using the LEWITE performance test as H\&R was deemed impractical in an online setting, as participants could easily look up vocabulary definitions online. This issue would likely persist in a laboratory setting due to the accessibility of mobile devices.
Therefore, a set of multiple-choice verbal analogy questions -- previously used in the SAT analogy test \citep{turney2006} -- was adopted instead\footnote{Example question: Which pair of words relates to each other like ”liter” to ”volume”? A) day - night; B) mile - distance; C) decade - century; D) friction - heat E) part - whole. Answer "B)" is correct here because the first word represents a unit of measurement for the second as "liter" does for "volume".}. This task, paired with a per-item time limit, was selected because it met multiple criteria for comparability with the original study.  
First, identifying analogies targets similar cognitive abilities as defining vocabulary, combining semantic, verbal, and logical reasoning skills.
Second, analogies are not easily searchable online, especially under time pressure. 
Third, SAT questions feature a sufficient number of distractors (four) besides the correct response to reduce the likelihood of success through guessing, albeit not as effectively as LEWITE\footnote{The LEWITE scheme consists of a sentence with two gaps and seven to nine answer options. In the easiest case, guessing would therefore be successful in $1/7*1/6=1/42=2.38\%$ of attempts, assuming independence between answers for simplicity). In comparison, the success probability of guessing on an SAT item is $1/5 = 20\%$.}.
Finally, these analogies require advanced vocabulary and have been shown to be relatively complex \citep{church2017}. Therefore, they should be sufficiently challenging -- particularly in combination with a per-item time limit -- to be considered "difficult" in the sense  of H\&R. Also, they should allow for enough performance variance to create an informative ranking for the test-based payoff condition.
Like in H\&R, the performance test consisted of 20 items. From a pool of 337 non-validation SAT test items, 23 were randomly selected, translated to German, and checked for potential region-specific terms. Twenty items were then used as tasks for the performance test, while the remaining three served as sample questions (see Appendix \ref{instructions}).

Potential limitations of the alternative task format will be discussed in Section \ref{discussion}.




\subsection{Sample}

A between-subject experimental design was used to compare two experimental groups of one-hundred participants each, one using a probabilistic-outcome distribution lottery mechanism mimicking the one from H\&R and one incorporating a modified fixed-outcome distribution version (see \ref{replication} and \ref{adaptation}). The sample size was determined via power analysis using G* Power 3.1 \citep{faul2009}. Following the experimental design with the goal of comparing two independent groups, assuming a medium effect size $d = 0.4$ \citep{cohen2013}, and using the social science convention of $a = 0.05$ (two-tailed), 80\% power \citep{brysbart2019}, 100 subjects per group were needed for sufficiently powered results, resulting in a total sample size of two-hundred subjects ($N = 200$).

At the time of the study, the Prolific participant pool contained 4,475 individuals who matched the study specification of "first language: German". This criterion was deliberately selected to align with the original language of H\&R’s study and its subject pool. 
Each subject was allowed to participate in only one experimental condition. While H\&R’s sample was mainly comprised of student participants recruited from the university campus, the Prolific worker base is more heterogeneous in terms of socio-demographic characteristics \ref{tab:demographics}). Prolific is a well-established platform for recruiting participants for academic studies from whom a certain level of knowledge, cognitive abilities, and effort can be expected. Additionally, testing the robustness of H\&R’s results to a non-student sample should be insightful in itself.
Unlike H\&R’s original study, which conducted multiple smaller sessions, this experiment featured a single large group per treatment condition, altering the number of individuals against whom subjects form their relative self-assessment. The potential implications of this aspect are discussed in Section \ref{discussion}.


\subsection{Procedure}

\begin{figure}[H]
    \centering
    \includegraphics[scale=0.45]{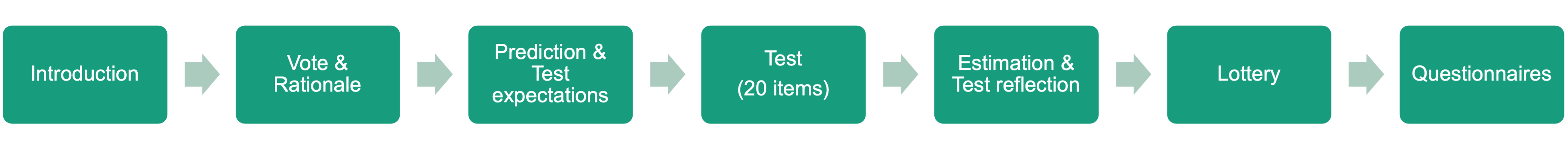}
    \caption{Overview of experimental procedure}
    \label{fig:procedure}
\end{figure}

An overview of the basic experimental procedure is provided in Figure \ref{fig:procedure}.

After being thanked in advance for their participation and providing consent for data processing, participants received the experimental instructions. A countdown of 90 seconds was set as minimum reading time before they could proceed to the next page. The instructions provided detailed information about the voting decision, the two bonus payment mechanism alternatives, and the subsequent experimental procedure. Next, extensive comprehension checks on key aspects of the experimental design were conducted to ensure proper understanding of the rules. Participants were only allowed to advance to the vote after correctly answering all comprehension checks. 

Similar to H\&R, subjects were then presented with three sample performance test items, which they could solve to familiarize themselves with the task type. As in H\&R, participants were informed that the test items could be categorized as difficult. On the subsequent page, the correct solution to the sample items were displayed alongside the participants' answers. Subjects were then informed that the vote would take place on the following page.

To cast their vote, participants simply clicked a radio button next to either "Test" or "Lottery". The explanations of the two payoff mechanisms were displayed above the selection buttons to ensure proper understanding. Beyond H\&R's procedure, after casting their vote, participants were asked to provide a brief rationale (at least 20 characters) for their choice to better understand their voting motives and gain additional insight into their decision-making process.

Next, exactly as in H\&R, subjects predicted their own score as well as the average score among all participants before proceeding to the performance test. Furthermore, they were asked how sure they were to have made the right choice in the vote, how difficult they would find it to change their decision, how important was doing well in the test to them, how difficult they thought the test would be and how well they thought they will do in it.

Unlike in H\&R, the test included a 30-second countdown per item to prevent participants from searching for answers online -- an issue that was not relevant at the time of the original study.

After completing all 20 test items, participants were shown their test score along with the correct solutions. They were subsequently asked to (re-)estimate the average test score among all participants and to indicate their satisfaction with their test results, how difficult they found the test, how sure they were now to have made the right choice in the vote and how difficult they would find it now to change it (all questions identical to those in H\&R). 
Subsequently, the lottery part commenced with participants selecting a "lucky number" between 1 and 6. The lottery design constituted the experimental treatment variation (described in detail in Sections \ref{replication} and \ref{adaptation}).

Following the lottery, participants completed a multi-part questionnaire that included three standardized scales, two price list tasks, and basic demographic questions.

Since social comparison aversion and low self-efficacy may be potential reasons discouraging individuals from voting for a performance-based payoff scheme test in favor of the lottery, the questionnaire contained a six-item short version of the Iowa–Netherlands Comparison Orientation Measure (INCOM) \citep{schneider2014,gibbons1999} to assess participants' inclination toward engaging in social comparison, as well as short-measure for general self-efficacy \citep{beierlein2013}, serving as controls. Also, the altruism sub-scale of the HEXACO-PI-R-100 \citep{lee2018} was included as a measure of individuals' prosocial tendencies. Afterward, participants completed the multiple price list format by \citet{dohmen2011} to elicit individual risk preferences as well as the analogously designed lottery task by \citet{gneezy2015} to measure individual attitudes towards ambiguity\footnote{Neither task was incentivized to prevent dilution or hedging of incentives from the main experiment. Prior research has shown that response behavior in these types of lotteries does not significantly differ between medium-to-high hypothetical incentives and small real incentives \citep{holt2002,gneezy2015}.}. While ambiguity aversion usually appears stable for same individual on different tasks, related literature emphasizes the importance of jointly assessing risk and ambiguity attitudes, as otherwise either may be overestimated \citep{krahnen2014,gneezy2015}. Additionally, the simple self-report measure of general risk preferences by \citet{dohmen2011} was included as a control for response consistency since it is more intuitive than the lottery format. Finally, subjects answered demographic questions regarding their age, gender, education level, student status and study major or professional occupation. 

To ensure attentiveness, two attention checks were embedded in the experiment -- one at the very beginning and another within the INCOM questionnaire. At the end of the study, participants were also asked directly whether they had paid attention and completed the experiment in one go. Their response had no effect on their compensation, allowing for voluntary disclosure of inattentiveness or carelessness.


\subsection{Replication Condition} \label{replication}

In the Replication condition, the analogous die roll from H\&R was emulated by a random generator draw. Subjects were asked to choose a "lucky number" between 1 to 6. After all 100 subjects had participated, a random number between 1 and 6 was drawn. If this number was even, all subjects who had chosen an even lucky number received the bonus payment, while those who had chosen an odd number won if an odd number was drawn.

Thus, subjects had a 50\% chance of receiving a bonus payment, independent of other subjects' choices, in contrast to the test condition, where an individual’s chance of winning partly depended on the performance of others. Consequently, this yielded a probabilistic-outcome distribution of 50 winners and 50 non-winner in expectation, meaning that among all 100 participants, 50\% would, on average, receive the bonus payment  (provided the majority vote favored the lottery payoff mechanism). However, the actual outcome distribution would deviate from an exact 50/50 split if subjects' number choices were non-uniformly distributed, just as could happen with outcomes in a traditional die roll.

\subsection{Adaptation Condition} \label{adaptation}

In the Adaptation condition, the lottery format was modified to ensure a fixed-outcome distribution, eliminating variance in the number of bonus recipients. Specifically, the lottery determined exactly 50 winners and 50 non-winners through random draw. As a result, the lottery-based payoff scheme becomes more comparable to the performance-based one, since -- unlike in the Replication condition -- the number of winners is predetermined, with no variance. Also, similar to the test condition, the chance for receiving a bonus payment becomes interdependent among subjects. Consequently, instead of being binomially distributed as in the Replication condition, the probability of receiving a bonus follows a hypergeometric distribution.

To maintain procedural parallelism with the Replication condition, and to avoid considerations related to a potential illusion of control \citep{langer1975,presson1996}, subjects in the Adaptation group were also asked to choose a "lucky number" from 1 to 6. Together with a subject's Prolific worker ID, the chosen number formed their unique "win code". After all 100 subjects had participated, 50 codes were drawn at random to determine which subject would receive a bonus payment (provided the majority vote favored the lottery payoff mechanism). If the majority vote favored the performance-based bonus payoff, conditions remained exactly the same as in the Replication condition.

\subsection{Incentive Scheme}
Subjects received a fixed payment of GBP 3.00 and were informed about the conditions for receiving a bonus payment in the instructions. The possibility of receiving a bonus payment was also prominently announced on the study's invitation page. The experiment was estimated to last approximately 25 minutes, based on test runs with student assistants, while H\&R's original experiment took around 30 minutes. This translates to an hourly rate GBP 7.20, which is categorized as "fair" by Prolific (the minimum required hourly reward is GBP 6.00). However, since -- in expectation -- half of all subjects would receive the bonus payment, the average hourly earnings increased to GBP 13.20, exceeding Prolific’s recommended rate of GBP 9.00 and making it competitive compared to other economic experiments.

Once all 100 subjects in a group had completed the experiment, a bulk mail was sent via the Prolific platform, announcing the the voting result and the median test score. The median test score automatically determined the cutoff for bonus payments under the performance-based scheme. Alternatively, if the lottery-based scheme was chosen, the worker IDs of those subjects receiving the bonus were listed. The bonus payments were credited to the respective participants' accounts shortly afterward.

\section{Results} \label{results}

A total of three-hundred individuals ($N=300$) participated in the experiment, completing it in an average of 23 minutes. Originally, the research plan aimed for 100 participants per treatment condition, totaling 200 participants. However, the two original sub-samples exhibited a statistically significant gender distribution imbalance. Coupled with an observed gender difference in voting tendencies -- 58.9\% of female participants voted for the test, compared to 75.2\% of male participants (Pearson $\chi^2(1) = 5.42, p = 0.020$) -- this created inference issues in assessing a potential treatment effect. To control for potential gender bias and ensure an unbiased inference of treatment effects without confounding factors, a second, gender-balanced sample of 100 participants was drawn for the Replication condition\footnote{The original Replication group comprised 30\% women and 70\% men, whereas, by chance, the gender distribution in the Adaptation group was exactly balanced at 50/50 (Pearson $\chi^2(1) = 7.56, p = 0.006$). Since Prolific offers the possibility to recruit samples with predefined demographic criteria, this feature was leveraged to enhance the comparability and of the results.}.

Although all participants affirmed to have paid sufficient attention and effort, 18 observations were excluded from the analysis: 10 from the Replication group and 8 from the Adaptation group. Nine failed at least one attention check, eight provided voting rationales that were nonsensical, arbitrary or inconsistent to their vote, and one explicitly stated an intent to bias the results. Thus, 182 subjects enter the subsequent analysis. 
The resulting sample had an average age of 35.9 years (Std. dev. = 11.8; median = 33.0), ranging from 19 to 74. This was noticeably higher than in H\&R (23 years), reflecting the broader demographic composition of Prolific compared to H\&R’s predominantly student-based sample. The sample was 51\% female. The most common fields of study and professions were "Business, Economics, Marketing, Sales, or Insurance' and 'Education, Cultural Studies, or Public Sector,' each representing close to 20\% of participants, followed by 'Information Technology and Natural Sciences' at just over 15\%. Appendix \ref{appendix} provides a detailed demographic overview.

\subsection{Choice of Vote}

Unlike in H\&R, the test payoff mechanism received the majority vote in both treatments, with 67 and 66 votes in the Replication and Adaptation condition, respectively. As shown by one-sided Binomial tests in Table \ref{tab:vote}, the voting share in favor of the test significantly exceeds the 50\% reference point in both groups.

\begin{table}[H]
    \centering
    \caption{Average Vote for Test, by Treatment}
    \begin{adjustbox}{max width=\textwidth}
    \begin{tabular}{l c c c }
         \hline
         \hline
          &  Mean &  Std. dev. & \makecell{Binomial test \\ p-value}\\
         \hline
         Replication  & 0.744 & 0.439 & 0.0000 \\
         Adaptation  & 0.717 & 0.453 & 0.0000 \\
         \hline
         \hline
    \end{tabular}
    \end{adjustbox}
    \label{tab:vote}
\end{table}

However, the distribution of voting shares does not differ significantly between the Replication and Adaptation groups (Pearson $\chi^2(1) = 0.17, p = 0.681$). Accordingly, the data do not support Hypothesis 1.

Whereas H\&R observed underplacement, the voting distributions in both the Adaptation and Replication conditions suggest overplacement. Thus, while the test received significantly more votes than the 39\% observed in H\&R ($p < 0.0001$ for one-sided Binomial tests in both groups) which means 'less underplacement' than in H\&R in a literal sense, this difference is clearly not attributable to the voting mechanisms' characteristics. Consequently, Hypothesis 2 is also not supported.

In the initial gender-unbalanced Replication sample, 65.5\% of participants voted for the test -- based on 90 observations that remained in the sample after applying the same exclusion criteria as in the other groups -- which would also have led to the rejection of the hypotheses (Pearson $chi2(1) = 0.8087, p = 0.369$). 
Since the lottery mechanism's design does not appear to alter voting outcomes between treatments, all available observations ($N = 272$) are pooled for the subsequent analysis in order to examine the factors and motivations underlying individual voting behavior. Additional treatment comparisons are provided in Table \ref{tab:treatment_comparison} in Appendix \ref{appendix}.

\subsection{Subjects’ Assessments} \label{predictions}


On average, subjects predicted their own performance to be nearly 14 correct answers out of 20, which is approximately two more than the average in H\&R's incentivized treatments (11.35). Correspondingly, their own performance prediction exceeded their prediction of the group's mean performance by nearly one point (see Table \ref{tab:predictions}), indicating a statistically significant difference in expected performances (Wilcoxon signed-rank test: $|z| = 5.64, p < 0.0001$). 
Meanwhile, subjects' predictions of the group’s mean performance closely aligned with the corresponding prediction in H\&R (13.09).
Using the ratio between the two predictions -- H\&R's "better" indicator -- subjects, on average, expected their own performance to be 8.8\% above the group mean. This contrasts with the direction of H\&R's findings, where subjects in the \textit{Money × Difficult} treatment reported a ratio of 0.88, indicating they expected to perform, on average, 12\% worse than their peers. 
 
\begin{table}[H]
    \centering
    \caption{Subjects Performance Predictions and Ratio Before the Test}
    \begin{adjustbox}{max width=\textwidth}
    \begin{tabular}{l c cc }
         \hline
         \hline
         & & \multicolumn{2}{c}{Pooled sample} \\\cline{3-4} 
         Variable & &  Mean &  Std. dev. \\
         \hline
         Predicted own performance & & 13.94 & 3.612 \\
         Predicted group performance & & 12.99 & 2.594  \\
         Better & & 1.088 & 0.269  \\
         \hline
         \hline
    \end{tabular}
    \end{adjustbox}
    \label{tab:predictions}
\end{table}


Overall, nearly two-thirds (63.97\%) of participants predicted outperforming the group average, which is a proportion significantly higher than 50\% (One-sided Binomial test: $p = 0.0000$) and substantially higher than in H\&R, where only 33\% of subjects made such predictions, peaking at 47\% in the \textit{Easy × No Money} treatment.

Examining the consistency between performance predictions and voting behavior (see Table \ref{tab:consistency_predictions_vote}), 87.4\% of subjects who predicted to outperform the group voted for the performance-dependent payoff mechanism (i.e., the test), compared to 40.8\% of those who predicted performing at or below the group average. This substantial and statistically significant difference (McNemar's $\chi^2(1) = 5.23, p = 0.0300$) underscores the connection between self-assessment and choice of compensation scheme.
Mathematically equivalent, 79.6\% of test voters expected to outperform the group average, compared to only 27.5\% of lottery voters. 

\begin{table}[H]
    \centering
    \caption{Consistency of Predictions and Voting Behavior}
    \begin{adjustbox}{max width=\textwidth}
    \begin{tabular}{l  c c  c }
         \hline
         \hline
            & \multirow{2}{*}{Test} & \multirow{2}{*}{Lottery} & \multirow{2}{*}{Total} \\
         Prediction & & & \\
         \hline
         Own $>$ Group  & 152 & 22 & 174 \\
         Own $\leq$ Group  & 40  & 58 & 98 \\
         \hline
         Total & 192 & 80 & 272 \\
         \hline
         \hline
    \end{tabular}
    \end{adjustbox}
    \label{tab:consistency_predictions_vote}
\end{table}


Accordingly, while the majority subjects' votes aligns with their self-assessments, notable deviations from equilibrium strategy behavior persist. From an equilibrium perspective, \textit{all} subjects expecting outperform the group should have chosen the test; yet, over 12\% opted for the lottery. More strikingly, more than 40\% of those anticipating their performance to be at or below the group mean still voted for the test. These discrepancies suggest that factors beyond self-assessment likely influenced voting behavior (see Section \ref{ratinoales}).

On average, participants solved 12.8 correctly (see Table \ref{tab:actual_post}), falling short of their own predicted performance by 1.1 points. This gap is notably larger than in H\&R's \textit{Difficult} treatments (approx. 7 correct answers) but smaller than in their Easy treatments (approx. 16 correct).

\begin{table}[H]
    \centering
    \caption{Subject Performance and Estimated Group Performance After the Test}
    \begin{adjustbox}{max width=\textwidth}
    \begin{tabular}{l c cc }
         \hline
         \hline
         & & \multicolumn{2}{c}{Pooled sample} \\\cline{3-4} 
         Variable & &  Mean &  Std. dev. \\
         \hline
         Actual own performance & & 12.84 & 3.551 \\
         Estimated group performance & & 12.52 & 2.594  \\
         \hline
         \hline
    \end{tabular}
    \end{adjustbox}
    \label{tab:actual_post}
\end{table}

\begin{figure}[H]
    \centering
    \includegraphics[width=0.8\linewidth]{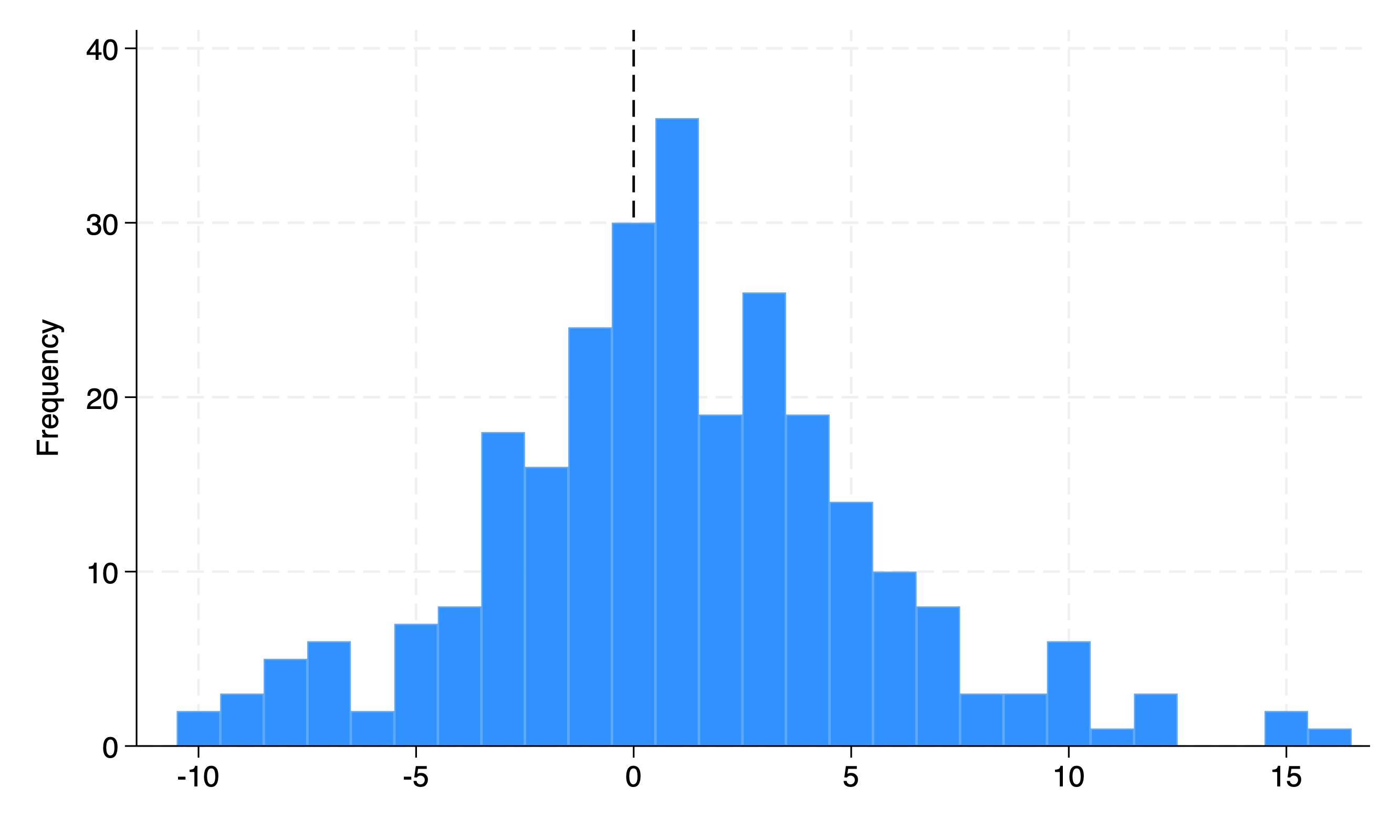}
    \caption{Frequency distribution of differences between predicted and actual performance}
    \label{fig:estim_accuracy}
\end{figure}

Subjects' own performance predictions were significantly correlated with their actual performance (Spearman's $\rho = 0.2435, p = 0.0001$), indicating that individuals were quite good at assessing their performance, in tendency. As in H\&R, in a simple linear regression of actual performance on predicted performance yielded a positive and significant coefficient, though its magnitude was roughly one-third of that in H\&R (see Table \ref{tab:regression_actual} in Appendix \ref{appendix}). The coefficient of determination being small was attributed to large individual errors by H\&R.
A similar pattern can also be observed in this study, with subjects' estimates of their own scores differing significantly from their actual performance (Wilcoxon signed-rank test: $|z| = 4.01, p = 0.0001$). The discrepancy featured substantial variation (Std. dev. = 4.5 points) caused by instances of extreme overestimation (up to 16 points) and underestimation (up to 10 points), as can be seen in Figure \ref{fig:estim_accuracy}. Nevertheless, 30 participants (11\%) precisely predicted their own score, while one-third (33.1\%) of provided predictions within one point of their actual score. Overall, a tendency for overestimation can be observed, with more than half of subjects' (55.5\%) own performance predictions exceeding their actual performance. Simultaneously, one third (33.5\%) of subjects underestimated their own performance. 

Following test completion and learning their individual scores, participants revised their estimates of group performance downward by an average of 0.5 points -- a statistically significant difference from pre-test predictions (Wilcoxon signed-rank test: $|z| = 2.57, p = 0.0100$).
Despite this adjustment, the difference between subjects' own performance (now known with certainty) and their estimated group mean remained significant (Wilcoxon signed-rank test: $|z| = 2.87, p = 0.0041$), though the absolute gap narrowed. This pattern aligns with \citet[p. 511]{moore2008}'s prediction that individuals' "beliefs about their own scores should be a joint function of their prior expectations and the private signals they receive regarding their own performance (i.e., their experience taking the quiz)".
The observed downward revision suggests that subjects extrapolated their own underperformance onto their group estimate while maintaining the directional belief that they outperformed the group, albeit with a reduced margin. 
This adjustment may be explained by subjects placing greater weight on their own performance signal than on beliefs about peer performance, as noted by \citep{moore2007cain}.
Meanwhile, the proportion of participants estimating the group to have performed worse or equally to themselves (54.8\%) shifts significantly closer to those estimating the group to have performed better (45.2\%), compared to pre-test predictions (McNemar's $\chi^2(1) = 7.53, p = 0.0080$).

Item-wise comparisons of subjects' responses to pre- and post-test perception questions -- such as being sure about one's voting decision, importance of doing well in the test, expected task difficulty, and expectation of doing good -- between this study and H\&R can be found in Tables \ref{tab:summary_statistics_pre} and \ref{tab:summary_statistics_post} in Appendix \ref{appendix}. 
In the pre-test assessments, scores for all five items are higher in this study. While the level of certainty about having made the right voting decision ("sure") is only marginally higher than in H\&R, the differences in the other four measures are substantial and highly statistically significant. Most notably, perceived task difficulty and the stated importance of doing well appear particularly elevated, exceeding H\&R's values by 0.9 and 2.7 points, respectively, on a 7-point scale.
However, these differences must be interpreted with caution, as H\&R only report averages across their entire sample despite the likelihood that these expectations were influenced by at least one of their treatment manipulations. For instance, the actual task difficulty should affect both perceived task difficulty and subjects' performance expectations\footnote{Perceived task difficulty before the test was significantly correlated to subjects' expectation to do good in it (Spearman's $\rho = -0.3659, p < 0.0001$). Meanwhile, the expectation to do good was significantly correlated with their predicted score (Spearman's $\rho = 0.6296, p < 0.0001$) and, consequently, with the prediction ratio "better" (Spearman's $\rho = 0.5105, p < 0.0001$), but only weakly and insignificantly correlated with their actual performance (Spearman's $\rho = 0.1062, p = 0.0803$).}. Similarly, while real monetary incentives -- which apply to all participants in this study -- should affect importance placed on performing well and, potentially, both subjects' certainty level about their voting decision and the perceived ease to change it. The same considerations apply to post-task reflections.

After the test, subjects in this study reported slightly lower satisfaction with their performance compared to those in H\&R -- potentially due to performing well in the sample task but falling short in the actual test. Additionally, they were significantly less sure about having made the correct voting decision post-test than both subjects in H\&R (Two-sample t-tests, equal variances: $t = -2.47, p = 0.0139$) and themselves pre-test (Wilcoxon signed-rank test: $|z| =  3.11, p = 0.0018$), with a decline of approximately 0.4 points in both cases. This shift can likely be attributed to the signal received through learning their test results (see Sections \ref{pred_corr_vote} and \ref{ratinoales}), as quiz scores and post-test certainty level are significantly correlated (Spearman's $\rho = 0.3162, p = 0.0000$)\footnote{This correlation is clearly driven by test voters, as the relationship is particularly strong and highly significant among them (Spearman's $\rho = 0.5094, p < 0.0001$), whereas no such correlation is observed among lottery voters (Spearman's $\rho = 0.0071, p = 0.9499$). This pattern is intuitive: subjects who voted for the test and learned they performed well could -- in a vacuum -- conclude to have made a good decision, despite having no information on their relative standing yet.}. 
Furthermore, perceived task difficulty increased slightly (by 0.2 points on average) but significantly (Wilcoxon signed-rank test: $|z| = 3.35, p = 0.0008$) from pre-test expectations to post-test assessments, indicating that subjects found the task somewhat more challenging than anticipated. This pattern contrasts with H\&R, where perceived difficulty did not increase post-test. In absolute terms, subjects in H\&R rated their task as 1.2 points easier than those in this study. Notably, the fact that subjects, on average, still estimated their own performance to be superior to that of their peers despite perceiving the task as more difficult than expected contrasts with the findings of \citet{healy2007}.
Meanwhile, subjects’ perceived difficulty in changing their vote remained virtually unchanged from pre- to post-task, whereas in H\&R, it increased by half a point. 
However, it is important to note that in H\&R, the die roll took place immediately after the test with the winning numbers known at the time. As a result, subjects were aware of whether they would receive a bonus if the lottery won the vote when completing the post-test questionnaire. This additional information may have influenced their assessment of both their level of being sure about their vote and their willingness to change it. In contrast, in this study, the lottery outcome was revealed only after all subjects had participated. Lacking this key information may explain why subjects' certainty level about the voting decision was lower than in H\&R.


\subsection{Predictors and Correlates of vote} \label{pred_corr_vote}

Following the analysis from H\&R, a logit regression of the vote on "better" -- defined as the ratio of predicted own performance to predicted group performance, serving as implicit measure of self-assessment relative to the group -- reveals a positive relationship between the probability of voting for the test and assessing one's own performance as superior to the group's average (see Table \ref{tab:logit_vote_ratio}). This relationship is highly significant, with a marginal effect (at the mean) of 0.769 ($p < 0.0001$), indicating a 76.9 percentage-point increase in the likelihood of voting for the test if a a predicts their own performance to be superior to the group's average.

\begin{table}[H]
    \centering
    \caption{Logit regression of Vote for Test Over Ratio of Performance Predictions}
    \label{tab:logit_vote_ratio}
    \begin{adjustbox}{max width=\textwidth}
    \begin{threeparttable}
    \begin{tabular}{l c c c c c c }
         \hline
         \hline
          & & pseudo-$R^{2}$ & & $\chi^{2}$ & & Prob $ > \chi^{2}$ \\\cline{3-7}
          & & 0.2209 & & 37.61 & & 0.0000\\
         \midrule
         Vote for Test & & Coefficient & & $z$ & & $P>z$ \\
         \hline
         Better   & & 5.0265 & &  6.13  & & 0.000  \\
         Constant  & & -4.3622 & & -5.23  & & 0.000 \\
         \hline
         \hline
    \end{tabular}
    \begin{tablenotes}
    \small
      \item \textit{Note}: \textit{Better}: Individual ratio between predicted own performance and predicted group. Regression model estimated using robust standard errors. n = 272.
    \end{tablenotes}
    \end{threeparttable}
    \end{adjustbox}
\end{table}

Consistent with H\&R, a logit regression of test voting behavior using predicted own performance and predicted group performance as independent variables yields highly significant coefficients of the expected signs (see Table \ref{tab:logit_vote_performance}). Predicted own performance positively affects the likelihood of voting for the test (marginal effect at the mean: $0.066, p < 0.0001$), whereas predicted group performance has a negative effect (marginal effect at the mean: $-0.050, p < 0.0001$). Thus, a one-point increase in predicted own performance is associated with a 6.6 percentage-point increase in the likelihood of voting for the test, while a one-point increase in predicted group performance corresponds to a 5.0 percentage-point decrease. 

\begin{table}[H]
    \centering
    \caption{Logit regression of Vote for Test Over Performance Predictions}
    \label{tab:logit_vote_performance}
    \begin{adjustbox}{max width=\textwidth}
    \begin{threeparttable}
    \begin{tabular}{l c c c c c c }
         \hline
         \hline
          & & pseudo-$R^{2}$ & & $\chi^{2}$ & & Prob $ > \chi^{2}$ \\\cline{3-7}
         & & 0.2500 & & 45.23 & & 0.0000\\
         \midrule
         Vote for Test & & Coefficient & & $z$ & & $P>z$ \\
         \hline
         Predicted own performance  & & 0.4550 & & 6.72 & &  0.000\\
         Predicted group performance  & & -0.3472 & & -4.40 & &  0.000\\
         Constant  & & -0.7580 & & -0.86 & &  0.390 \\
         \hline
         \hline
    \end{tabular}
    \begin{tablenotes}
    \small
      \item \textit{Note:} Regression model estimated using robust standard errors. n = 272.
    \end{tablenotes}
    \end{threeparttable}
    \end{adjustbox}
\end{table}

\paragraph{Signals}

Furthermore, it appears reasonable to assume that subjects incorporated information from solving the sample questions into their self-assessment and voting decision. Additionally, this information may have influenced their assessment of the group’s performance, despite the absence of direct feedback on competitors’ performance.

On average, subjects correctly answered 2.28 out of 3 sample questions (Std. dev. = 0.843). However, as they had unlimited time and no restrictions against external assistance, cheating cannot be ruled out. In fact, sample question performance is strongly and significantly correlated with performance in the main task (Spearman's $\rho = 0.498, p < 0.0001$). Additionally, higher sample question performance is associated with with higher own performance predictions (Spearman's $\rho = 0.417, p < 0.0001$)\footnote{Similarly, a significant correlation of sample question performance with both the pre-task control question on how good subjects think they will do in the test (Spearman's $\rho = 0.209, p = 0.0005$) and the ratio of own and group predictions ("better") can be observed (Spearman's $\rho = 0.184, p = 0.0024$). Moreover, sample question performance is negatively correlated with perceived task difficulty before the test (Spearman's $\rho = -0.1826, p = 0.0025$).}. 
In line with these correlations, subjects who performed better on the sample questions were significantly more likely to vote for the test as the payoff mechanism. The proportion of test voters increased progressively, from 10.0\% among those who solved no sample questions correctly, to 43.6\% with one correct, 70.5\% with two correct, and 83.0\% with all three correct responses (Pearson $\chi^2(3) =  41.33, p < 0.0001$)\footnote{The share of test voters increases significantly from each number of correctly solved sample questions to the next.}. 
Similarly, when extending the logit regression of voting behavior on the two predictions by adding sample performance as a predictor (see Table \ref{tab:logit_vote_example_perform}), the marginal effect (at the mean) of sample performance on test voting probability is 0.105 ($p < 0.0001$). This inclusion reduces the marginal effect of own performance prediction while increasing the overall explanatory power of the model.

\begin{table}[H]
    \centering
    \caption{Logit regression of Vote for Test Over Performance Predictions and Sample Performance}
    \label{tab:logit_vote_example_perform}
    \begin{adjustbox}{max width=\textwidth}
    \begin{threeparttable}
    \begin{tabular}{l c c c c c c }
         \hline
         \hline
          & & pseudo-$R^{2}$ & & $\chi^{2}$ & & Prob $ > \chi^{2}$ \\\cline{3-7}
         & & 0.2916 & & 47.72 & & 0.0000\\
         \midrule
         Vote for Test & & Coefficient & & $z$ & & $P>z$ \\
         \hline
         Predicted own performance  & & 0.3996 & & 5.73 & &  0.000\\
         Predicted group performance  & & -0.3825 & & -4.55 & &  0.000\\
         Sample question performance  & & 0.7745 & & 3.35 & &  0.001\\
         Constant  & & -1.2589 & & -1.39 & &  0.164 \\
         \hline
         \hline
    \end{tabular}
    \begin{tablenotes}
    \small
      \item \textit{Note:} Regression model estimated using robust standard errors. n = 272.
    \end{tablenotes}
    \end{threeparttable}
    \end{adjustbox}
\end{table}

Interestingly, higher sample question performance was also associated with increased group performance predictions (Spearman's $\rho = 0.2685, p < 0.0001$). This finding somewhat contradicts the argument by H\&R that individuals fail to adjust their expectations of others’ performance to newly acquired task-related information.

As it could be expected, individuals who voted for the test reported a higher confidence higher expectation to do well in it, averaging 5.2 on a 7-point scale, compared to 4.3 among lottery voters (Two-sample Mann-Whitney U-Test: $|z| = 5.35, p < 0.0001$). Test voters also assigned greater importance to performing well than lottery voters (6.1 vs. 5.6 on a 7-point scale), though the latter still reported relatively high importance (Two-sample Mann-Whitney U-Test: $|z| = 2.69, p = 0.0073$).
Perceived task difficulty did not differ significantly before the test (Two-sample Mann–Whitney U-test: $|z| = 1.73, p = 0.0832$). However, lottery voters perceived the task as slightly more difficult beforehand and downgraded their assessment afterward (5.6 to 5.4, Wilcoxon signed-rank test: $|z| = 1.17, p = 0.2434$), while test voters initially perceived the task as less difficult but upgraded their assessment the task and upgraded perceived difficulty (5.4 to 5.8, Wilcoxon signed-rank test: $|z| = 4.67, p = 0.0000$). The level of being sure about their voting decision and the perceived difficulty to change did not differ by vote.

\paragraph{Questionnaire Controls}

Table \ref{tab:questionnaire_comparison_vote} presents the mean scores for the additional questionnaire controls introduced in this study (beyond those in H\&R) across the full sample, as well as differentiated by voting decision.

\begin{table}[H]
    \centering
    \caption{Questionnaire Summary Statistics}
    \label{tab:questionnaire_comparison_vote}
    \begin{adjustbox}{max width=\textwidth}
    \begin{threeparttable}
    \begin{tabular}{l c cc c cc c cc}
         \hline
         \hline
         & & & & &  \multicolumn{5}{c}{Vote}\\\cline{6-10}
         & & \multicolumn{2}{c}{Pooled sample} & &  \multicolumn{2}{c}{Test} & &  \multicolumn{2}{c}{Lottery}\\\cline{3-4} \cline{6-7} \cline{9-10}
         Variable & &  Mean &  Std. dev. & & Mean &  Std. dev. & & Mean & Std. dev. \\
         \hline
         INCOM & & 3.47 & 0.74 & & 3.44 & 0.73 & & 3.55 & 0.78 \\
         GSE & & 3.87 & 0.68 & &  3.82 & 0.72 & &  3.89 & 0.66 \\
         Altruism & & 3.79 & 0.69 & & 3.77 & 0.67 & &  3.84 & 0.74 \\
         Risk (MPL) & & 15.26 & 5.72 & & 15.19 & 5.47 & &  15.45 & 6.31 \\
         Risk (11-point) & & 5.51 & 2.47 & & 5.43 & 2.50 & & 5.69 & 2.42 \\
         Ambiguity & &  12.24 & 5.54 & & 12.47 & 5.61 & & 11.70 & 5.35 \\
         \hline
         \hline
    \end{tabular}
    \begin{tablenotes}
    \small
      \item \textit{Note:} \textit{INCOM}: Short scale of Iowa–Netherlands Comparison Orientation Measure (5-point scale); \textit{GSE}: General self-efficacy (5-point scale); \textit{Altruism}: Altruism facet from HEXACO-100 (5-point scale); \textit{Risk (MPL)}: Multiple price list format for measuring risk preferences; \textit{Risk (11-point)}: General willingness to take risk (11-point scale); \textit{Ambiguity}: Multiple price list format for measuring ambiguity aversion; N = 272; Test n = 192; Lottery n = 80.
    \end{tablenotes}
    \end{threeparttable}
    \end{adjustbox}
\end{table}

Participants exhibited a moderate tendency for social comparison, averaging 3.5 on a 5-point scale. While one might intuitively expect individuals with a stronger inclination for social comparison to favor the test, the reported value is slightly -- but insignificantly-higher among lottery voters (Two-sample Mann–Whitney U-test: $|z| =  0.95, p = 0.3448$). The same pattern holds for general self-efficacy. Although medium to high levels of self-efficacy were observed across the sample, the expectation that individuals with higher self-efficacy would be more inclined to enter a performance-based competition is not reflected in the data (Two-sample Mann–Whitney U-test: $|z| = 0.94, p = 0.3509$). Similarly, participants demonstrated medium to strong levels of altruism, yet no statistical difference was found between test and lottery voters (Two-sample Mann-Whitney U-test: $|z| = 0.73, p = 0.4684$), providing no evidence that altruism drives the decision to vote for the lottery.
H\&R themselves acknowledged ambiguity aversion as a potential influence on voting behavior, given that participants were choosing between a lottery with known probabilities and a test with an ambiguous probability of success. Using the elicitation price list from \citet{gneezy2015}, the average switching point from a lottery with known probabilities to ambiguous lottery occurred between rows twelve and thirteen, suggesting a general tendency toward ambiguity aversion, with participants demanding an ambiguity premium of approximately one-fourth of the potential winning amount. However, switching behavior did not differ significantly between test and lottery voters (Two-sample Mann-Whitney U-test: $|z| = 1.03, p = 0.3039$), and the distributions of switching points\footnote{To circumvent multiple switches between columns, the procedure suggested by \citet{andersen2006} has been adopted in which only one row has to be designated for a switch from the lottery to the certainty equivalent.} were virtually identical (Two-sample Kolmogorov–Smirnov test: $D = 0.1000, p = 0.587$).
Self-reported risk attitudes -- measured using both an 11-point general risk scale and a multiple price list format \citep{dohmen2011} -- located, on average, around the midpoint of the respective scales. Participants' willingness to take risks was relatively consistent between the two measures (Spearman's $\rho = 0.289, p < 0.0001$), and may overall be classified as approximately risk-neutral, though both measures exhibited standard deviations of around 40\%. Neither measure revealed significant differences between test and lottery voters (Two-sample Mann-Whitney U-tests: General risk question: $|z| =  0.61, p = 0.5424$; Multiple price list: $|z| = 0.52, p = 0.6013$).
Meanwhile, participants' risk and ambiguity attitudes were significantly correlated (Spearman's $\rho = 0.135, p = 0.0257$).

In summary, contrary to initial expectations, none of the additional latent constructs measured -- namely social comparison tendency, general self-efficacy, altruism, risk preferences, or ambiguity attitudes -- differed significantly between test and lottery voters (nor could any treatment differences be observed; see Table \ref{tab:treatment_comparison} in Appendix \ref{appendix}).


Regarding demographic controls, no significant age differences were found between test and lottery voters (Two-sample Mann-Whitney U-test: $|z| = 0.66, p = 0.5101$). 
Meanwhile, as previously noted, voting behavior appears to differ significantly by gender: while 62.5\% of female subjects voted for the test, 75.2\% of male subjects did (Pearson $\chi^2(1) = 6.77, p = 0.009$).  
Additionally, 78.4\% of students voted for the test compared to 66.8\% of non-students,  a borderline statistically significant difference (Pearson $\chi^2(1) = 3.83, p = 0.050$). This discrepancy could be attributed to students ascribing more intelligence to themselves or being more accustomed to test-taking, despite lacking information about the cognitive abilities and student status of other participants. Moreover, a marginally insignificant difference in voting behavior was observed between subjects with a degree in higher education (i.e., current or former students) and those without one: 74.9\% of those with a degree voted for the test, compared to 63.8\% of those without (Pearson $\chi^2(1) = 3.79, p = 0.052$).
Interestingly, actual test performance did not differ significantly between men and women (Two-sample Mann-Whitney U-test: $|z| = 0.22, p = 0.8247$) or between students and non-students (Two-sample Mann-Whitney U-test: $|z| = 1.11, p = 0.2671$), which would justify the preference of the former for the test\footnote{Similarly, no significant differences were found between these groups in sample question performance (Two-sample Mann–Whitney U-tests: Gender: $|z| = 1.70, p = 0.0896$; Student: $|z| = 0.49, p = 0.6352$).}.


For completeness, Appendix \ref{appendix} presents the results of a logistic regression model (Tables \ref{logit_coefficients} and \ref{logit_margins}) examining the likelihood of voting for the test, based on all previously discussed variables -- including performance predictions, sample question performance, demographics, pre-test assessments, and questionnaire controls. These results largely support and reinforce the findings from non-parametric analysis. Performance predictions and signals from sample items remain strong and highly significant predictors of voting the test across all model specifications, while gender and student status also appear to be relevant factors, as discussed. A small yet significant negative effect of social comparison orientation emerges, indicating that a stronger tendency to engage in social comparison is associated with a lower likelihood of voting for the test. Contrary to initial expectations, this may suggest that, at least in this context, participants who are more inclined to compare themselves to others tend avoid competition, possibly influenced by the perceived difficulty of the task. However, this interpretation remains speculative.

\paragraph{Calibration}

According to the equilibrium strategy outlined by H\&R, a rational decision maker should vote for the test if they assess their skill to be above the payoff-relevant cutoff – in this case the median performance -- and should vote for the lottery otherwise.
Under this assumption, a subject can be classified as accurately calibrated if they:
$(i.)$ voted for the test and their performance was in the upper half of their group’s performance distribution, or -- equivalently -- 
$(ii.)$ voted for the lottery and their performance was in the lower half of their group’s performance distribution\footnote{In case a participant's performance was exactly at the median, they were also classified as accurately calibrated.}. 

Based on this criterion, 68.4\% of participants made an accurate voting decision, representing a significant deviation from a fully accurately calibrated population (One-sided Binomial test: $p = 0.0000$). A breakdown of calibration accuracy across multiple sub-groups of the sample is presented in Table \ref{tab:calibration}.

\begin{table}[H]
    \centering
    \caption{Calibration Accuracy, by Sub-groups}
    \begin{tabular}{l l c c c c}
        \hline
        \hline
        & & & Fraction & & Pearson $\chi^2$-Test \\ \cline{3-6}
        \multicolumn{2}{l}{\textbf{Sub-group}} & & & & \\
        \hline
        \multicolumn{2}{l}{\textbf{Vote}} & & & & \\        
        & Test & & 0.641 & & \multirow{2}{*}{$\chi^2(1) = 5.63, p = 0.018$} \\
        & Lottery & & 0.788 & &  \\
        \hline
        \multicolumn{2}{l}{\textbf{Gender}} & & & & \\        
        & Men & & 0.704 & &  \multirow{2}{*}{$\chi^2(1) = 0.65, p = 0.422$} \\
        & Women & & 0.658 & &  \\
        \hline
        \multicolumn{2}{l}{\textbf{Student status}} & & & & \\
        & Students & & 0.693 & & \multirow{2}{*}{$\chi^2(1) =  0.05, p = 0.818$}  \\
        & Non-students & & 0.679 & &  \\
        \hline
        \multicolumn{2}{l}{\textbf{Degree in higher education}} & & & & \\
        & Yes & & 0.731 & & \multirow{2}{*}{$\chi^2(1) = 4.37, p = 0.037$} \\
        & No & & 0.610 & &  \\
        \hline
        \multicolumn{2}{l}{\textbf{Sample question performance}} & & & & \\
        & 0 correct & & 0.900 & & \multirow{4}{*}{$\chi^2(3) = 5.78, p = 0.123$} \\
        & 1 correct & & 0.590 & &  \\
        & 2 correct & & 0.636 & &  \\
        & 3 correct & & 0.726 & &  \\
        \hline
        \hline
    \end{tabular}
    \label{tab:calibration}
\end{table}

Lottery voters were more accurately calibrated than test voters, meaning their vote was more consistent with their subsequent performance.
Among those who voted for the test, 64.1\% actually performed better than the median, whereas 78.8\% of those who voted for the lottery performed worse than the median of their respective group. This statistically significant difference aligns with the tendency of participants to overestimate rather than underestimate their own performance in absolute terms in the predictions. 

Unlike for voting behavior, the proportion of accurately calibrated participants did not differ significantly by gender or student status. Also, there was no significant difference in age distributions between well-calibrated and inaccurately calibrated participants (mean age of well-calibrated: 33.4, Std. dev. = 12.7; mean age of inaccurately calibrated: 36.3, Std. dev. = 13.2; Two-sample Mann–Whitney test: $ |z| = 1.14, p = 0.2541$).
However, participants with a university degree were significantly better calibrated -- by approximately 12 percentage-points -- than those without a degree.

Additionally, calibration accuracy followed a slight U-shaped distribution across different sample performance levels, with participants who performed either very well (all questions correct) or very poorly (none correct) being more accurately calibrated than those with intermediate performance. 

As it may be expected, accurately calibrated subjects were significantly more sure about having made the correct choice in the vote after the test than inaccurately calibrated subjects (Two-sample Mann-Whitney U-test: $|z| = 4.22, p < 0.0001$), reporting averages of 5.2 vs. 4.2 on a 7-point scale. Before the test, however, certainty levels did not differ significantly (Two-sample Mann-Whitney U-test: $|z| = 1.42, p = 0.1564$). This suggests that learning their individual test scores provided participants with a clearer sense of whether they had assessed their own abilities correctly, even though they still lacked information on their competitors' performance.
Fittingly, inaccurately calibrated subjects reported a significantly higher perceived task difficulty after completing the test compared to accurately well-calibrated subjects (6.0 vs. 5.5 on a 7-point scale;, Two-sample Mann-Whitney U-test: $|z| = 3.85, p = 0.0001$), but not before (Two-sample Mann-Whitney U-test: $|z| = 0.90, p = 0.3706$). They were also significantly less satisfied with their test outcome than well-calibrated subjects (3.3 vs. 4.5, Two-sample Mann-Whitney U-test: $|z| = 4.61, p = 0.0000$).


\subsection{Self-reported Voting Motives} \label{ratinoales}

\begin{figure}[H]
    \centering
    \includegraphics[width=0.9\linewidth]{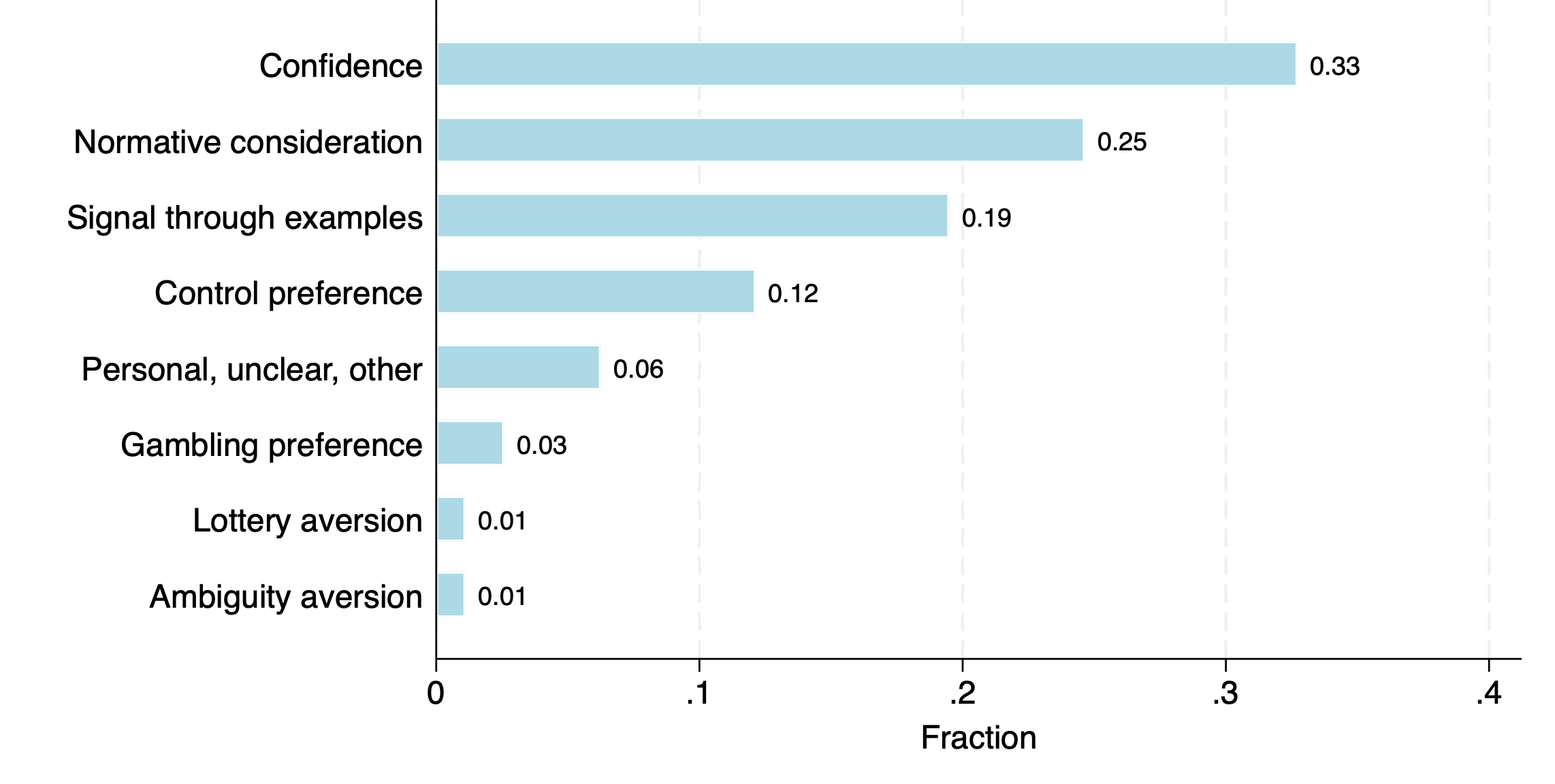}
    \caption{Voting rationales, Self-reported by Subjects {(N = 272)}}
    \label{rationale_sample}
\end{figure}

Immediately after casting their vote in the experiment, subjects were asked to provide a short rationale for their choice. This allows for a deeper understanding of whether individuals followed the equilibrium strategy outlined by H\&R or had alternative -- monetary or non-monetary -- motives underlying their vote that may also explain inaccurate calibration. 
This section offers an exploratory-descriptive analysis of these self-reported rationales, grouping them into inductively generated thematic categories reflecting latent meanings. \citep{braun2006}. The relative frequencies of these motive categories across the full sample are displayed in Figure \ref{rationale_sample}. Example statements for each of the main rationales can be found in Table \ref{tab:categorization}\footnote{The complete list of subjects' self-reported rationales and their categorization can be obtained from \url{https://osf.io/ucx53}.}.
It is important to note that these self-reported motives are not necessarily mutually exclusive (e.g., confidence may be altered by signals through sample question outcomes). Therefore, the observed proportions should be interpreted with caution and regarded as indicative tendencies rather than precise measurements. Consequently, statistical tests and cross-comparisons are deliberately avoided.

\begin{figure}[H]
    \centering
    \includegraphics[width=1\linewidth]{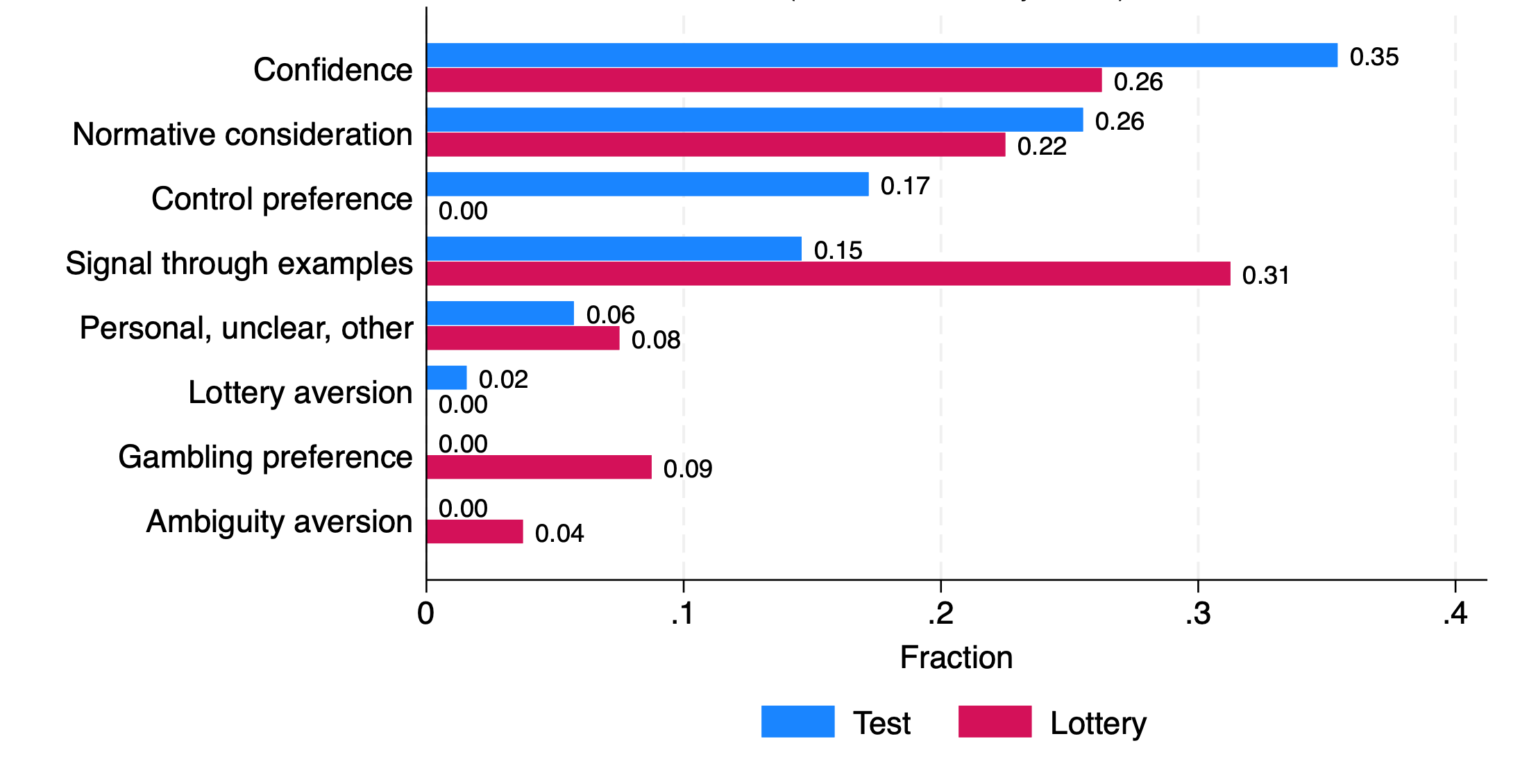}
    \caption{\small Voting Rationales by Choice in Vote, Self-reported by Subjects; Test: n = 192; Lottery: n = 80}
    \label{rationales_vote}
\end{figure}

According to their responses, approximately one-third of subjects’ voting decisions were driven by confidence in their abilities and performance expectations. 
Differentiating between high and low confidence (see Figure \ref{rationales_vote}) -- high confidence resulting in test votes and low confidence in lottery votes -- around 35\% of test voters and 26\% of lottery voters attributed their decision to confidence.  
A preference for control -- in the sense of the potential bonus payment depending on personal performance rather than luck \citep[see][]{owens2014,benoit2022} -- was cited by 17\% of test voters (equivalent to 12\% of all participants), while -- logically -- no lottery voters reported this rational. 
Beyond these two rationales, that were already discussed in the literature (see Appendix \ref{validity}), approximately one-fifth of participants reported basing their decision on their sample question performance, with this factor being more prevalent among lottery voters (31\%) than test voters (15\%). These patterns aligns with the significant differences in voting behavior by sample question outcomes noted in Section \ref{pred_corr_vote}, where preference for the test progressively increased with better sample performance. The possibility that sample question results could serve as a decision-making signal was not considered in H\&R.
Additionally, normative beliefs played a substantial role, with meritocratic beliefs reported by 26\% of test voters and preferences for equal chances cited by 22\% of lottery voters. As suggested by \citet{krawczyk2012}, notions of fairness were invoked to justify both perspectives \citep[see also][]{cappelen2007pluralism}: some test voters viewed performance-based allocation as the fairest option, even if they did not expect to receive a bonus themselves. Meanwhile, lottery voters considered it fair if all subjects had equal statistical chances of receiving the bonus. This motive category appears otherwise unrecognized in the related literature on performance-based versus lottery-based selection mechanisms. The cited lottery rational aligns with this study's arguments made in Section \ref{hypotheses}. 
Notably, with 36\%, normative beliefs represent the most common rational among inaccurately calibrated subjects (see Figure \ref{rationales_accuracy} in Appendix \ref{appendix}), re-emphasizing the role of non-monetary motives leading to -- apparently -- inaccurate self-assessments in the vote from an equilibrium strategy perspective.

Other motives, such as ambiguity aversion and preferences for or against gambling, were mentioned only occasionally\footnote{Logically, not all rationales apply to both voting outcomes: a preference for control and lottery aversion were only reported by test voters, while ambiguity aversion -- a possible motive for choosing the lottery discussed by H\&R -- and a preference for gambling were exclusive to lottery voters.}. 

While confidence remains the most frequently cited rationale, the data clearly highlight the prominence of alternative non-monetary considerations, such as control preferences and normative beliefs. Moreover, the prevalence of confidence as the voting rationale may still be overstated, not only due to the non-exclusive nature of the response categorization but also because confidence itself is less clearly defined than other motives. For instance, some of the self-reported low confidence among lottery voters may stem from ambiguity aversion without participants explicitly stating it or even being aware of it. Unlike well-articulated beliefs such as fairness considerations or control preferences, confidence represents a more latent sentiment that sometimes serves as a catch-all explanation or a diagnosis of exclusion for various intentional or subconscious behavioral influences \citep{merkle2011}.

\begin{table}[H]
    \centering
    \caption{Example Statements for Self-reported Voting Rationaes, by categorization}
    \begin{tabular}{l l p{10.5cm}}
        \hline
        \hline
        \multicolumn{2}{l}{\textbf{Rationale}} & \textbf{Example}\\
        \hline
        \multicolumn{2}{l}{\textbf{Confidence}} & \\        
        & High & "I believe that I [can/will] deliver an above-average performance and therefore have the best chance of receiving the bonus payment with the performance variant." \\
        & Low & "Not enough confidence in my own abilities to definitely finish in the top 50." \\
        \hline
        \multicolumn{2}{l}{\textbf{Preference for control}} & \\        
        & & "I chose ‘Test’ because I prefer the bonus to depend on my own ability and performance rather than chance/luck." \\
        \hline
        \multicolumn{2}{l}{\textbf{Sample question signal}} & \\
        & Positive & "I got all the sample tasks correct. If the upcoming tasks are similarly easy/difficult, I have a good chance of finishing in the top 50."  \\
        & Negative & "I only got 1 out of 3 [sample questions] correct, logically the probabilities are higher if I take the lottery." \\
        \hline
        \multicolumn{2}{l}{\textbf{Normative beliefs}} & \\
        & Principle of merit & "Those who have better knowledge should be rewarded accordingly. I don't expect to be one of the 50 people to listen, but I think the principle is fair." \\
        & Equal chances & "People should have opportunities regardless of their performance. Performance as a factor is only fair if everyone has the same starting conditions. This is never the case in reality."\\
        \hline
        \hline
    \end{tabular}
    \label{tab:categorization}
\end{table}



\section{Discussion} \label{discussion}

This study aimed to replicate the findings by H\&R and examine their robustness against an alternative lottery mechanism design. By translating their original experiment to an online environment and extending it with standardized questionnaires and a self-reported rationale statement, it may be inferred whether miscalibrated voting distributions arise from inaccurate self-assessments or other factors when individuals choose between a performance-based and a random device-based payoff scheme.

The results reveal voting distributions more consistent with traditional notions of overplacement than with the underplacement pattern found in H\&R, as -- clearly -- no underplacement is observed in any group. The alternative fixed-outcome distribution lottery mechanism is not found to impact voting outcomes. Regardless of treatment, approximately nearly three-fourth of participants preferred the performance-test over the lottery for determining the bonus payment allocation. Correspondingly, participants tended to overestimate their absolute test performance rather than underestimate it.

Generally, the same predictors of voting behavior identified in H\&R -- predicted own performance, predicted group performance, and the expectation of outperforming the group average ("better") -- are also observed here, with predicted own performance being the most influential. Beyond H\&R’s findings, sample question performance, gender, and student status are additionally found to be significantly related to voting behavior. Notably, sample performance serves as a strong signal for decision-making, with preference for the test over the lottery increasing progressively with the number of correctly solved sample questions -- an effect not explicitly acknowledged in H\&R. Meanwhile, social comparison tendencies, general self-efficacy, altruism, risk attitude, ambiguity aversion, and other demographic characteristics show no significant relationship with voting behavior.

An exploratory analysis of participants’ self-reported voting rationales reveals four primary drivers of voting behavior. Besides confidence -- whether high or low -- participants frequently cite their sample question performance as a basis for their decision, reinforcing the correlation patterns observed in this study. While this approach is not strictly rational from an objective perspective -- since it neglects the competitive aspect of performance assessment \citep[][]{hoelzl2005,alicke2005,kruger2008} -- it appears to play a substantial role in participants’ voting decisions. 
Additionally, many test voters express a preference for control, indicating a desire to actively influence their payoff basis through the test, despite its inherent ambiguity due to the bonus payment being contingent on others' performance. 
Finally, a substantial proportion of participants state normative beliefs as their main voting motive. This applies to both choice alternatives, with test voters referencing meritocratic preferences and lottery voters citing equality of opportunity. These self-reported rationales, at least anecdotally, align with the argument put forth in this study, even though its hypotheses are not explicitly supported. 
Since the latter two rationales -- control preferences and normative beliefs -- are not grounded in payoff-maximizing considerations, they contradict the equilibrium strategy postulated by H\&R.


Overall, participants’ behavior and self-assessments appear reasonably consistent, with the calibration of above- and below-median performers opting for the test and lottery, respectively, being relatively accurate at close to 70\%. While using sample performance as a proxy for test performance is not objectively rational -- since it conflates absolute and relative evaluations -- it is a subjectively understandable heuristic, as it serves as the only available reference point for participants. 
While the present sample is more demographically diverse than H\&R’s original sample, this does not seem to have significant implications -- except, perhaps, that students show a greater inclination toward voting for the test than non-students. However, this trend is not clearly reflected in comparisons with H\&R’s results, as their sample includes a larger proportion of students.

Limitations of this study concern the comparability between the performance test used in the original study and the alternative task employed in this replication. 
Using the task from H\&R's 2005 experiment -- a vocabulary-based knowledge test -- in its original form would have been impractical under modern technological conditions, both in online and lab settings, due to the prevalence of mobile devices giving participants the opportunity of searching for answers online. Even with modifications such as a timer, the informative value of the results would have been limited. 
Consequently, this study replaces it with an alternative format -- logical analogy selection under time pressure -- designed to engage similar cognitive abilities. Since anticipated task difficulty, which is difficult to assess in isolation, plays a crucial role in expected self-assessment miscalibration, any divergence between this study’s results and those of the original should be interpreted with caution. Based on average performance, this task’s difficulty falls between H\&R’s "difficult" and "easy" tasks. Thus, it may be assumed that participants did not perceive it as genuinely “difficult” (H\&R’s \textit{Difficult} groups solved an average of 7 items). However, participants in this study rated the task as more difficult than subjects in H\&R even before taking the test, and significantly adjusted their stated difficulty perceptions upward post-test -- despite performing well on both sample questions and the test itself. Some explicitly emphasized task difficulty in their self-reported rationales.

A key issue is the absence of an objective measure of difficulty, as H\&R differentiate task difficulty only in relative terms. Moreover, they report only the average perceived task difficulty across all participants, despite having a design that explicitly distinguishes between two difficulty levels.
In the current study, a logit regression of voting behavior on pre-test difficulty assessments yields a non-significant coefficient that is barely different from zero ($\beta1 = -0.102, |z| = 0.63, p = 0.531$; M.E.: $-.0219, p=0.529$), suggesting that factors other than task difficulty play a more central role in voting decisions -- consistent with the study’s findings.
H\&R report a similar non-effect in all groups except the \textit{Difficult × Money} condition, where they find a counterintuitive positive and significant coefficient, implying an increasingly strong preference for the test the more difficult it is perceived under monetary incentives. However, H\&R do not further discuss potential reasons for this effect.
Furthermore, the task in this study offered a ten-fold higher chance of success through guessing. However, in principle, this should not affect relative overconfidence. To counterbalance this and prevent participants from searching for answers online, time pressure was introduced to increase the perceived difficulty of the task. 

Apart from the different task, two other minor factors distinguish this replication from the original experiment (see also Table \ref{tab:comparison} in Appendix \ref{appendix}), primarily for operational reasons.
First, the increased physical distance between participants in an online environment may have influenced decision-making. Better-than-average effects are usually found to be stronger when people compare themselves to an abstract group rather than to a specific known individual \citep{moore2007cain}, and tend to weaken with personal contact or when comparison targets become more individuated and concrete \citep{alicke1995}. This distinction may partially apply to the contrast between centralized lab settings and remote participation, potentially leading to differences in voting behavior — even though no specific information about comparison targets was provided in either case, and direct interaction between subjects was absent in the lab as well. However, online experiments have generally been found to provide "adequate and reliable" \citep[p. 100]{arechar2018} alternatives to lab-based studies, with both approaches yielding consistent and robust results \citep{prisse2022}.
Second, this study featured a significantly larger reference group -- 100 participants in each group -- compared to the original experiment, where subjects were placed in smaller groups of "at least nine". While a larger reference group should not impact a subject’s relative self-assessment from an equilibrium perspective, empirical findings suggest that an increased number of competitors can reduce competitive motivation \citep{garcia2009}. However, this should have promoted lottery votes rather than test votes, which was not the case in this experiment.
Consequently, the reasons for the observed differences in voting behavior between H\&R and this study remain largely unclear at this stage.

As in H\&R, the task in this study primarily assessed linguistic competencies, combined with elements of logical reasoning. Future research could further reinforce these findings by incorporating alternative test formats, such as mathematical problems or matrix-based tasks, as these may elicit different miscalibration dynamics across academic domains \citep{erickson2015}.
Furthermore, this study, like H\&R, does not cover motivational and status-based \citep[see e.g.,][]{alicke2005,anderson2012,brown2012,benabou2002} or social \citep[see e.g.,][]{proeger2014,tenney2019,taylor1988} accounts of overconfidence. These perspectives are effectively abstracted from the analysis and could provide valuable avenues for future research.

Replicating the classic experiment by H\&R, this study contributes to ongoing methodological discussions in experimental overconfidence research on the validity of using choices between performance-based and lottery-based payoffs as behavioral measures of overconfidence, as well as the broader conceptual validity of the overconfidence paradigm \citep[see e.g.,][]{benoit2011,merkle2011,owens2014}. Specifically, it seeks to gain a more comprehensive understanding of the factors contributing to misplacement in performance self-assessments beyond overconfidence and to explore alternative explanations for apparent overconfidence in overplacement data, thereby clarifying the scope of phenomena encompassed within the term "overconfidence".


\clearpage
\bibliographystyle{apalike}
\bibliography{btae}

\appendix

\clearpage
\section{Discussion on Overconfidence paradigm validity} \label{validity}


The connections between many of the phenomena discussed in Section \ref{introduction} and overconfidence have largely been established through argumentation and axiomatic reasoning rather than empirical evidence, as note by \citet{merkle2011}. Consequently, the apparent robustness of overconfident relative self-assessments as a stable psychological or behavioral pattern -- whether in the early assumption that individuals are generally overconfident \citep{debondt1995,alicke1995} or in the more nuanced perspective that individuals tend to overplace themselves on easy tasks while underplacing on difficult ones \citep{moore2008,kruger2008,grieco2009} -- may be questioned \citep{benoit2011,benoit2014,owens2014}. More specifically, it has been argued that "better-than-average" data may not be able to demonstrate overconfidence \citep{benoit2011}, and that inferring overconfident beliefs from choice behavior likely leads to overestimation of the actual extent of overconfidence \citep{owens2014}. 

In particular, \citet{benoit2011} challenge the axiomatically established presumption in psychology and economics that a rational population should exhibit neither overplacement nor underplacement. They argue that the statement \textit{"most people cannot be better than the median"} does not logically imply that \textit{"most people cannot rationally rate themselves above the median"} (p. 1592). Individuals may have valid reasons to believe they will perform in the upper half of a population. Therefore, misplacement in self-assessments is \textit{"unproblematic if [it]] can arise from a population whose beliefs are generated within a rationalizing model"} (p. 1594). According to their framework, this holds for almost any distribution of placement relative to the median. 
Thus, what appears to be overplacement may actually be consistent with rational Bayesian updating, given that individuals lack perfect knowledge of their own abilities and the overall distribution of skills in a population \citep{benoit2011,benoit2014,moore2008}. Under conditions of perfect information, only 50\% of individuals could, in fact, rationally place themselves in the top half of a given population.

With imperfect information, however, for a population to be truly considered overconfident, substantially more than 50\% would need to rate themselves above the median. Mathematically, \citet{benoit2011} derive that up to twice the proportion of any given percentile of a population may rationally rate themselves within that percentile (e.g., 60\% believing they are in the top 30\%) and propose this as a threshold for inferring overconfidence from overplacement data. Consequently, practically an entire population could rationally expect to be in its top half.

Accordingly, ranking experiments that rely on a single reference point -- typically the median -- and employ a dichotomous categorization of success to infer miscalibration in relative abilities generally reveal only 'apparent', rather than 'true', overconfidence or underconfidence. Many experimental findings of overplacement may therefore be median-rationalizable within \citet{benoit2011}'s theoretical framework, as demonstrated by \citet{benoit2014} in their reassessment of results from \citet{hoelzl2005} and \citet{moore2007cain}. This distinction is crucial because true overconfidence can lead to the negative real-world consequences outlined in Section \ref{introduction}, whereas apparent overconfidence is unlikely to have such effects \citep{benoit2014}.


Methodologically, \citet{benoit2011} criticize commonly used incentive-compatible probability elicitation measures, such as quadratic scoring rules, for often being unintuitive and failing to penalize incorrect estimations. Additionally, key aspects of individuals' subjective understanding of the skill or ability in questions -- such as their interpretation of "average" (whether as mean, median, or mode), response scales, and the signaling structure shaping their beliefs -- are typically unobservable, compromising both internal validity within studies and comparability across studies\footnote{For example, in \citet{svenson1981}'s prominent experiment on driving skill ranking -- besides it being unincentivized -- participants lacked objective criteria for accurately assessing their driving ability due to limited available information. Applying \cite{benoit2011}'s criteria, the reported proportion of better-than-median assessments suggests that the Swedish population’s responses can be median-rationalized, whereas the American population cannot. Consequently, Svenson identifies 'true' overconfidence in the American population but only 'apparent' overconfidence in the Swedish.} \citep{benoit2011,benoit2014}. Therefore, \citet{benoit2011} call for experimental designs that collect more detailed information on the strength of subjects’ beliefs beyond rankings relative to the median -- for instance, by asking participants to indicate their perceived likelihood of placing within specific deciles -- to be able to actually study observe overconfidence. Expanding on this, \citet{benoit2014} advocate for a more nuanced interpretation of individuals' self-assessments relative to the median. Subjects favoring placement-based rewards should not necessarily be interpreted as these individuals believing they are top-half performers with certainty -- which would indicate true overconfidence -- but rather as them assuming to have at least a 50\% chance of placing in the top half. This belief, they contend, is consistent with rational expectations that could be held by nearly every individual in a population.
In two experiments involving easy quizzes with complete information on prior performances, \citet{benoit2014} find overplacement data that are not rationalizable for a population of expected payoff maximizers under \citet{benoit2011}'s framework, and, therefore, appear to display true overconfidence. 
Their explanation of individuals extrapolating good absolute performance to above average relative performance while neglecting competitors' performance goes along with findings from prior research \citep[e.g.,][]{kruger1999,alicke2005,kruger2008,moore2007small}, and matches correlational patterns observed in this paper (see Section \ref{results}).

Several studies argue that overplacement should be considered an information or statistical bias rather than a confidence or behavioral bias, stemming from incomplete information on the respective reference group \citep{healy2007,moore2007small,blavatskyy2009}, or imperfect Bayesian updating to information about others \citep{moore2008,grieco2009,grossman2012}. 
Symptomatically, the tendency to rate oneself as above average on easy tasks and below average on difficult tasks becomes stronger if an individual receives information about their performance and is attenuated if they receive information about a reference group's performance \citep{moore2007small}. Moreover, incomplete information on a task’s difficulty or the distribution of relevant skills within the reference population may lead to either overconfident or underconfident beliefs \citep{healy2007,blavatskyy2009}. Especially, simultaneous inference about task difficulty and relative performance may lead to mixed and strongly context-dependent evidence on misplacement in either direction \citep{benoit2014,grieco2009,clark2009,moore2008}.
Therefore, the statistical impossibility of more than half a population assessing themselves in its top half does not necessarily indicate an underlying judgment bias, i.e., overconfident beliefs \citep{healy2007}.

Beyond the discussion on thresholds for justifiably inferring overconfidence from overplacement data and the role of information availability in shaping miscalibrated self-assessments, there may simply be additional non-monetary motives influencing individuals' placement behavior, which are difficult to disentangle from truely overconfident beliefs in discrete choice experiments. 
If overplacement in choice behavior is rooted in (true) overconfidence, i.e., a psychological misperceptions or errors in beliefs, better information could lead to adjusted behavior. Conversely, if overplacement is driven by alternative motives -- such as control preferences, self-enhancement, image concerns, or attitudes toward risk and ambiguity -- behavior is unlikely to change \citep{benoit2014,owens2014,benoit2022}.

For instance, \citet{heath1991} argue that humans exhibit a distinct preference for betting on themselves over a random device -- even when the latter offers an equal or higher success probability -- provided they feel competent in the given context. Beyond that, \citet{goodie2007} observe individuals to display a general preference for control across various domains.
\citet{owens2014} show that individuals sacrifice up to 15\% of expected earnings to bet on themselves — a "control premium"\footnote{"Control premium" is commonly used as an umbrella term that encompasses multiple reasons why individuals prefer to bet on themselves. These include a general preference for autonomy in decision-making, greater enjoyment of betting on oneself, the desire to signal competence, or a broader inclination toward self-reliance in wagering \citep{owens2014,benoit2022}.} that cannot be fully explained by beliefs. Instead of offering participants the choice between betting on their own performance and a random device -- as in H\&R -- they let subjects choose between betting on themselves or on someone else. Their results provide an estimate of more than half (57\%) of the observed self-reliance beyond 50\% stemming from a preference for control (64.9\% chose to bet on themselves overall), leaving the remainder attributable to overconfident beliefs \citep{owens2014}. Using these reference values to reinterpret H\&R's results, out of the 63 percent self-reliance in the \textit{Easy x Money} treatment, only 5.5 percentage-points of the deviation from 50 percent could be attributed to overconfidence, while the remaining 7.5 percentage-points would be explained by a preference for control. However, it must be noted that a desire for control likely would not explain underplacement data, apart from individuals exhibiting a (strong) negative control premium. \citet{owens2014} do neither explicitly comment on the \textit{Difficult x Money} condition from H\&R nor underplacement in general. 
\citet{benoit2022} replicate \citet{owens2014}'s experimental design with an adapted treatment condition in which participants choose between betting on their performance in one task versus another, ensuring individuals bet on themselves either way to mitigate potential control distortions related to ambiguity aversion towards another person's abilities. They also compare betting on one's own performance to betting on a random device with an identical probability of success. Their results suggest that at least 68\% of the apparent overconfidence observed in participants' choices between betting on their own performance and a random device (14.2 percentage-points in total) stemming from control-related preferences. When applying their adapted mechanism to mitigate control preferences, \citet{benoit2022} still observe a small but statistically significant degree of overplacement (54\% of participants expect to rank in the top 50\%), indicating the true overconfidence.

\citet{merkle2011} directly address Benoît \& Dubra's critique through two experiments that meet their criteria, by eliciting participants' complete belief distributions regarding their relative placement within a population across multiple domains and tasks -- capturing probability estimates for each decile or quartile rather than relying on point estimates -- to distinguish between rational information processing and (true) overconfidence as competing explanations for overplacement. Findings reveal considerable overplacement across several domains, while displaying underplacement for high specialization abilities (e.g., programming skills) and low-probability events (e.g., likelihood of suffering a heart attack), with probability estimates aligning with prior research based on point estimates. Rational information processing is rejected for four out of six domains, while results for the remaining two are mixed.
Consistent with \citet{benoit2014}, Merkle \& Weber find that overplacement in aggregated belief distributions is especially driven by low-skilled individuals strongly underperforming relative to already poor self-assessments, despite receiving negative signals.

Ultimately, \citet[p.262]{merkle2011} reaffirm that \textit{"overconfidence is not just an artifact of psychological experiments, but seems present in many real-life situations where considerable stakes are involved"}. While acknowledging the theoretical validity of Benoît and Dubra’s critique, they argue that its practical consequences may be limited, as overplacement persists even after accounting for rational adjustments. Accordingly, they conclude that \textit{"there is no need to discard previous literature on this premise"} (p. 271), though they support incorporating the discussed methodological refinements into experimental designs \citep{merkle2011}. 
\citet{benoit2014} oppose this by stating that, despite some evidence withstanding the criticism by \cite{benoit2011}, the actual scope and significance of overplacement in human judgment and decision-making remains unclear, limited number of studies employing rigorous experimental designs. Furthermore, they reemphasize that \textit{"in any case, it is important to realize that the degree of (true) overconfidence may not be well measured by the fraction of people who rank themselves as above average"} (p. 321).

\clearpage
\section{Supplementary Tables and Figures}    \label{appendix}

\begin{table}[H]
    \centering
    \caption{Demographic Statistics}
    \label{tab:demographics}
    \begin{adjustbox}{max width=\textwidth}
    \begin{threeparttable}
    \begin{tabular}{l c c c c c c c c}
         \hline
         \hline
         & & Total & & \makecell{Replication \\ (unbalanced)} & &  Adaptation & & \makecell{ Replication\\ (balanced)} \\ \cline{3-9} 
         Variable & & & & & & & &   \\
         \hline
         \textbf{Number of observations} & & 272 & & 90 & & 92 & & 90\\
         \textbf{Age} & &  34.3 & & 31.2 & & 34.1 & & 37.7 \\
         (SD) & &  (11.2) & & (9.4) & &  (11.6) &  & (11.9) \\
         \textbf{Female} & & 44.1 & &  30.0 & &  50.0  & & 52.2 \\
         \textbf{Student} & & 32.4 & & 38.9 & & 37.0 & & 21.1 \\
         \textbf{Study major or professional field} & & & & & &   \\
         Business, Economics, Marketing, Personnel, Sales, or Insurance & & 19.5 & &  18.9 & & 18.5 & & 21.1\\
         Design, Communications, or Media & & 4.0 & & 2.2 & & 6.5 & & 3.3\\
         Education, Cultural Studies or Public Sector & & 15.44 & & 7.8 & & 17.4 & & 21.1\\
         Engineering, Manufacturing, Construction, or Agriculture & & 10.3 & & 8.9 & & 14.1 & & 7.8\\
         Information Technology and Natural Sciences & & 18.4 & & 24.4 & & 15.2 & & 15.6 \\
         Medicine, Psychology, Health, or Social Care & & 8.5 & & 10.0 & & 7.6 & & 7.8\\
         Social Studies, Journalism, or Politics & & 4.0 & & 5.6 & & 5.4 & & 1.11\\
         Transport, Logistics, Retail, or Wholesale & & 5.5 & & 6.7 & & 6.5 & & 3.33 \\
         Homemaker, Unemployed, Retired, Pupil, or not specified & & 14.0 & & 14.4 & & 8.7 & & 18.9 \\
        \hline
         \hline
    \end{tabular}
    \begin{tablenotes}
    \small
      \item \textit{Note:} Averages reported for Age. Relative frequencies reported for all other variables.
    \end{tablenotes}
    \end{threeparttable}
    \end{adjustbox}
\end{table}

\begin{table}[H]
    \centering
    \begin{adjustbox}{max width=\textwidth}
    \begin{threeparttable}
    \caption{Subjects' Self- and Group Assessments Before and After the Test \& Questionnaire Control Summary Statistics, by Treatment}
    \begin{tabular}{l c cc c cc c cc}
         \hline
         \hline
         & & \multicolumn{2}{c}{Replication} & &  \multicolumn{2}{c}{Adaptation} & &  \multicolumn{2}{c}{Mann-Whitney U}\\\cline{3-4} \cline{6-7} \cline{9-10}
         Variable & &  Mean &  Std. Dev. & & Mean &  Std. Dev. & & z & p-value\\
         \hline
         \textbf{Before} & & & & & & & & \\
         Predicted own performance & & 13.54 & 3.20 & & 14.28 & 3.60 & & 1.77 & 0.0766\\
         Predicted group performance & & 12.27 & 2.40 & & 13.47 & 2.27 & & 3.18 & 0.0014 \\
         Better & & 1.118 & 0.25 & & 1.079 & 0.28 & & -0.32 & 0.7530 \\
         \hline
         \textbf{Before} & & & & & & & & \\
         Sample performance & & 2.36 & 0.85 & & 2.41 & 0.74 & & 0.18 & 0.8593\\
         Test performance & & 13.6 & 3.1 & & 13.1 & 3.3 & & -0.83 & 0.4074 \\
         \textbf{After} & & & & & & & & & \\
         Estimated group performance & & 12.6 & 2.6 & & 12.6 & 2.7 & & -0.13 & 0.8988\\
         Estimation accuracy & & -0.1 & 3.5 & & 1.2 & 4.3 & & 1.93 & 0.0530\\
         \hline
         \textbf{Questionnaire Controls} & & & & & & & & \\
         INCOM & & 3.56 & 0.66 & & 3.54 & 0.69 & & 0.22 & 0.8297 \\
         GSE & & 3.84 & 0.77 &  & 3.91 & 0.57 & & 0.03 & 0.9810\\
         Altruism & &  3.77 & 0.72 &  &  3.84 & 0.70 &  &  -0.79 & 0.4308\\
         Risk (MPL) & & 15.35 & 6.43 &  & 15.39 & 5.63 &  & -0.22 & 0.8269 \\
         Risk (11-point) & & 6.13 & 2.47 &  & 5.58 & 2.33 &  & 1.489 &  0.1368\\
         Ambiguity & & 12.58 & 5.86 &  & 12.16 & 5.43 &  & 0.416 & 0.6788\\
         \hline
         \hline
    \end{tabular}
    \label{tab:treatment_comparison}
    \begin{tablenotes}
    \small
      \item \textit{Note:} \textit{INCOM}: Short scale of Iowa–Netherlands Comparison Orientation Measure (5-point scale); \textit{GSE}: General self-efficacy (5-point scale); \textit{Altruism}: Altruism facet from HEXACO-100 (5-point scale); \textit{Risk (MPL)}: Multiple price list format for measuring risk preferences; \textit{Risk (11-point)}: General willingness to take risk (11-point scale); \textit{Ambiguity}: Multiple price list format for measuring ambiguity aversion. "Mann-Whitney U" reports results for Two-sample Mann–Whitney U-tests between treatments. Replication n = 90; Adaptation n = 92.
    \end{tablenotes}
    \end{threeparttable}
    \end{adjustbox}
\end{table}

\begin{table}[H]
    \centering
    \caption{Regression of Actual Performance over Predicted Own Performance}
    \label{tab:regression_actual}
    \begin{adjustbox}{max width=\textwidth}
    \begin{threeparttable}
    \begin{tabular}{l c c c c c c }
         \hline
         \hline
          & & $R^{2}$ & & $F$ & & Prob $ > F$ \\\cline{3-7}
          & &  0.0501 & & 13.66 & & 0.0003 \\
         \midrule
         Actual own performance & & Coefficient & & $t$ & & $P>t$ \\
         \hline
         Predicted own performance  & & 0.2201 (0.0595) & & 3.70 & & 0.000 \\
         Constant  & & 9.7702 (0.8420) & & 11.60 & &  0.000 \\
         \hline
         \hline
    \end{tabular}
    \begin{tablenotes}
    \small
      \item \textit{Note:} Refers to Table 16 in H\&R (Technical Appendix).
    \end{tablenotes}
    \end{threeparttable}
    \end{adjustbox}    
\end{table}

\begin{table}[H]
    \centering
    \begin{adjustbox}{max width=\textwidth}
    \begin{threeparttable}
    \caption{Summary Statistics, Before the Test}
    \begin{tabular}{l c cc c cc c cc}
         \hline
         \hline
         & & \multicolumn{2}{c}{Full sample} & &  \multicolumn{2}{c}{H\&R} & &  \multicolumn{2}{c}{Two-sample t-test}\\\cline{3-4} \cline{6-7} \cline{9-10}
         Variable & &  Mean &  Std. Error & & Mean &  Std. Error & & t & p-value \\
         \hline
         Predicted own performance & & 13.94 & 0.22 & & 12.51 & 0.30 & & 3.85 & 0.0001 \\
         Predicted group performance & & 12.99 & 0.16 &  & 13.37 & 0.24 & & -1.36 & 0.1751 \\
         Sure & & 5.29 & 0.08 &  & 5.16 & 0.14 &  & 0.85 & 0.3964\\
         Difficult to change & & 4.23 & 0.09 &  & 3.66 & 0.15 &  & 3.33 & 0.0009 \\
         Important & & 5.96 & 1.32 &  & 3.24 & 0.16 &  & 17.54 & 0.0000\\
         Difficulty test & & 5.46 & 0.06 &  & 4.58 & 0.10 &  & 7.62 & 0.0000\\
         Good in test & & 4.91 & 0.07 &  & 4.29 & 0.10 & &  4.92 & 0.0000\\
         \hline
         \hline
    \end{tabular}
    \label{tab:summary_statistics_pre}
    \begin{tablenotes}
    \small
      \item \textit{Note:} \textit{sure}: "How sure are you to have made the right decision in the vote?" (7 = very sure); \textit{difficult to change}: "How difficult would you find it to change your decision in the vote?" (7 = very difficult); \textit{important}: "How important is doing well in the test to you?" (7 = very important); \textit{difficulty test}: "How difficult do you think the test will be?" (7 = very difficult); \textit{good in test}: "How good do you think you will be in the test?" (7 = very difficult); N = 272; H\&R n = 134. Since H\&R report standard errors of the mean instead of standard deviations, this is adopted for comparability. Right column gives results from two-sided Two-sample t-tests with equal variances.
    \end{tablenotes}
    \end{threeparttable}
    \end{adjustbox}
\end{table}

\begin{table}[H]
    \centering
    \begin{adjustbox}{max width=\textwidth}
    \begin{threeparttable}
    \caption{Summary Statistics, After the Test}
    \begin{tabular}{l c cc c cc c cc}
         \hline
         \hline
         & & \multicolumn{2}{c}{Full sample} & &  \multicolumn{2}{c}{H\&R} & &  \multicolumn{2}{c}{Two-sample t-test}\\\cline{3-4} \cline{6-7} \cline{9-10}
         Variable & &  Mean &  Std. Error & & Mean &  Std. Error & & t & p-value\\
         \hline
         Actual own performance & & 12.84 & 0.22 & & 11.83 & 0.45 & & 2.29 & 0.0223 \\
         Estimated group performance & & 12.52 & 0.17 &  & 13.43 & 0.30 & & -2.86 & 0.0045 \\
         Satisfied & & 4.11 & 0.11 &  & 4.34 & 0.18 &  & -1.12 & 0.2624\\
         Sure after & & 4.88 & 0.10 &  & 5.33 & 0.15 & & -2.47 & 0.0139 \\
         Difficult to change after & & 4.42 & 0.11 &  & 4.15 & 0.18 &  & 1.31 & 0.1899 \\
         Difficulty test after & & 5.66 & 0.06 &  & 4.48 & 0.14 &  & 8.82 & 0.0000\\
         \hline
         \hline
    \end{tabular}
    \label{tab:summary_statistics_post}
    \begin{tablenotes}
    \small
      \item \textit{Note:} \textit{satisfied}: "How satisfied are you with your result in the test?" (7 = very sure); \textit{sure after}: "How sure are you now to have made the right decision in the vote?" (7 = very difficult); \textit{difficult to change after}: "How difficult would you find it now to change your decision in the vote?" (7 = very important); \textit{difficulty test after}: "How difficult do you think the test was?" (7 = very difficult); N = 272; H\&R n = 134. Since H\&R report standard errors of the mean instead of standard deviations, this is adopted for comparability. Right column gives results from two-sided Two-sample t-tests with equal variances.
    \end{tablenotes}
    \end{threeparttable}
    \end{adjustbox}
\end{table}

\begin{figure}[H]
    \centering
    \caption{Voting Rationales by Accuracy in Calibration, Self-reported by Subjects}
    \label{rationales_accuracy}
    \includegraphics[width=1\linewidth]{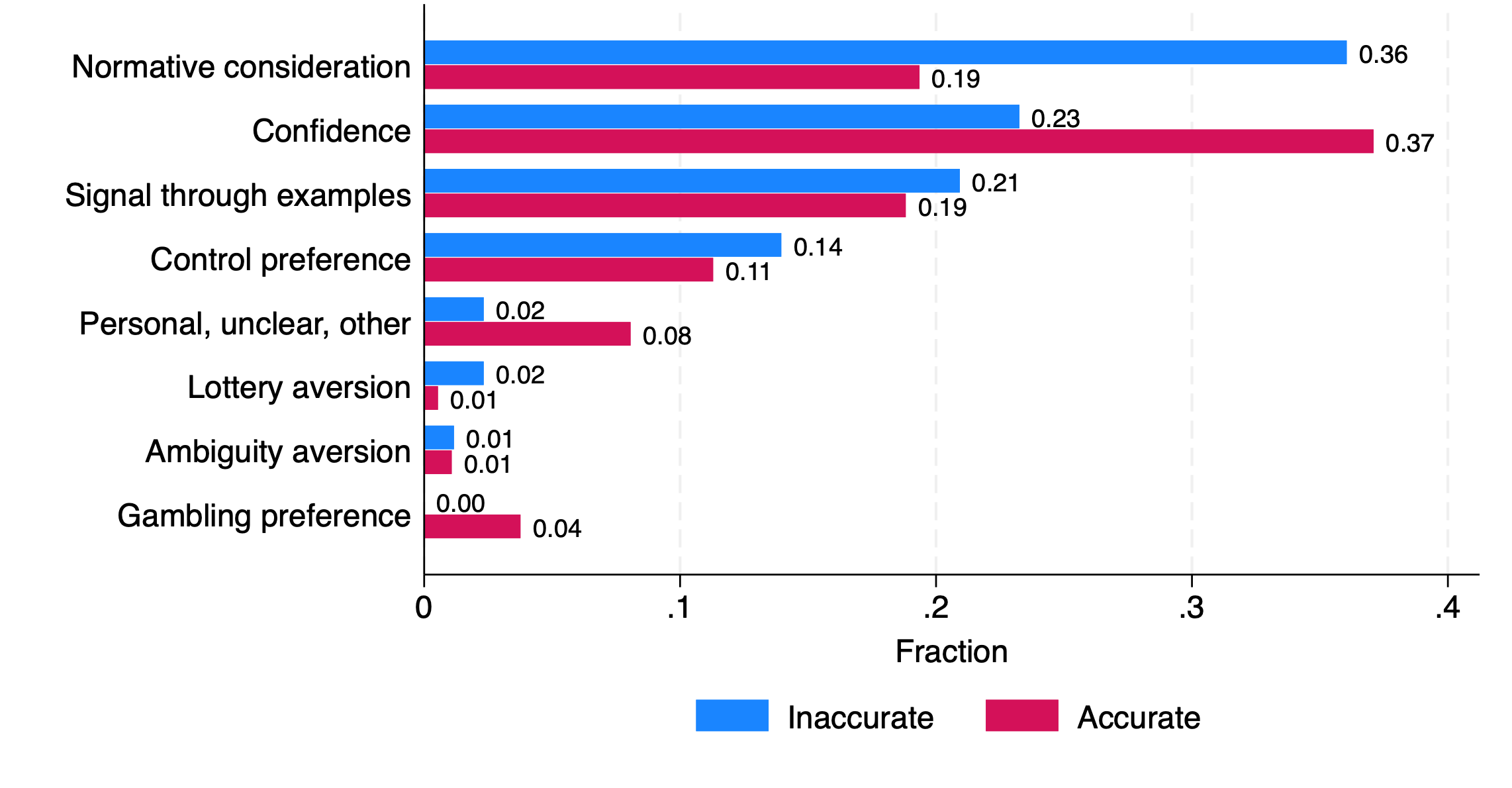}
    \caption*{\textit{Note}: Accurate: n = 186, Inaccurate: n = 86.}
\end{figure}

\clearpage

\begin{table}[H]
\centering 
\caption{Logit Regression for Voting Behavior -- Coefficients} 
\label{logit_coefficients}
\begin{threeparttable}
\begin{tabular}{l l c c c c c} 
\hline
\hline
 & & \multicolumn{5}{c}{Dependent variable: Likelihood of Vote for 'Test'} \\ \cmidrule(l){3-7} 
 & & (1) & (2) & (3) & (4) & (5) \\ 
\midrule
  \multicolumn{2}{l}{Intercept} & -1.259 &  -1.829 & -4.004 & 2.655 & -1.613 \\ 
  & & (0.905) & (0.925) & (1.796) & (1.640) & (2.296) \\
  &  &  &  &  &  &  \\ 
  \multicolumn{2}{l}{Own Performance Prediction} & 0.400 & 0.406 & 0.411 & 0.471 & 0.473 \\
  & & (0.070) & (0.072) & (0.090)  & (0.077) & (0.103) \\ 
 \multicolumn{2}{l}{Group Performance Prediction} & -0.382 & -0.384 &  -0.385 & -0.408 & -0.431 \\ 
  & & (0.084) & (0.084) & (.093) & (.082) & (0.101) \\ 
 \multicolumn{2}{l}{Sample Performance} & 0.775 & 1.041 & 0.858 &  0.685 & 1.148  \\ 
  & & (0.231) & (0.227) & (0.246) & (0.241) & (0.275) \\ 
  & & &  & &  &  \\
  \multicolumn{2}{l}{Age} & & 0.016 & & & 0.014\\ 
  & & & (0.024) & &  & (0.026) \\ 
  \multicolumn{2}{l}{Age$^{2}$} & & 0.002 & & & 0.002\\ 
  & & & (0.001) & &  & (0.001) \\ 
 \multicolumn{2}{l}{Female} &  & -1.372 &  &  &  -1.644 \\ 
  & & & (0.376) &  &   & (0.456) \\ 
 \multicolumn{2}{l}{Student} &  & 1.038 &  &  &  1.418\\ 
  & & & (0.444) &  &  & (0.490) \\ 
 \multicolumn{2}{l}{Degree in Higher Education} &  & 0.181 &  &   & 0.604 \\ 
  & & & (0.366) &  &  & (0.422) \\ 
  & & &  & &  &  \\
  \multicolumn{2}{l}{Sure to have made the right Vote} &  &  & -0.128 &  & -0.067 \\ 
  & & &  & (0.132) &  &  (0.159) \\ 
  \multicolumn{2}{l}{Difficulty to change Vote} &  &  & 0.044 &  & 0.021 \\ 
  & & &  & (0.115) &  &  (0.127) \\ 
  \multicolumn{2}{l}{Importance of doing well in Test} &  &  & 0.115 &  & 0.257 \\ 
  & & &  & (0.125) &  &  (0.166) \\ 
  \multicolumn{2}{l}{Expected Test Difficulty} &  &  & 0.303 &  & 0.354\\ 
  & & &  & (0.179) &  &  (0.227) \\ 
  \multicolumn{2}{l}{Expectation of doing well in Test} &  &  & 0.124 &  & 0.274  \\ 
  & & &  & (0.186) &  &  (0.203) \\ 
  & & &  & &  &  \\
  \multicolumn{2}{l}{Social Comparison Orientation (INCOM)} &  &  &  & -0.534 & -0.583 \\ 
  & & &  &   & (0.242) & (0.273) \\ 
 \multicolumn{2}{l}{General Self-Efficacy} &  &  & &  -0.407 & -0.622 \\ 
  & & &  &  & (0.261) & (0.353) \\ 
 \multicolumn{2}{l}{Altruism} &  &  &  & -0.168  & -0.168 \\ 
  & & &  &  & (0.280) & (0.371) \\ 
 \multicolumn{2}{l}{Risk Attitude} &  &  &  & -0.067  &  -0.121 \\ 
  & & &  &  & (0.075) & (0.092) \\   
 \multicolumn{2}{l}{Ambiguity Attitude} &  &  &  & 0.009 & 0.021\\ 
  & & &  &  &  (0.030) & (0.034) \\ 
\midrule
   \multicolumn{2}{l}{Wald $\chi^{2}(3)$} & 47.72 & 76.31 & 49.50 & 53.71 & 62.29 \\
   \multicolumn{2}{l}{Pseudo $R^{2}$} & 0.291 & 0.361 & 0.309 & 0.321 &  0.417 \\
   \multicolumn{2}{l}{N} & 272 & 272  & 272 & 272 & 272 \\
\hline
\hline
\end{tabular} 
\begin{tablenotes}
    \small
    \item \textit{Note:} Coefficients estimated using robust standard errors, standard errors in parentheses.     
    \item Model specifications: (1) Base model (see Table \ref{tab:logit_vote_example_perform}), (2) including demographics, (3) including pre-task expectations, (4) including questionnaire controls, (5) full model.
    Only the single-item 11-point risk measure by \citet{dohmen2011} is included for simplicity. Study major and professional field not included due to arbitrary categorization.
\end{tablenotes}
\end{threeparttable}
\end{table}

\begin{table}[H]
\centering 
  \caption{Logit Regression for Voting Behavior -- Marginal effects} 
  \label{logit_margins}
\begin{threeparttable}
\begin{tabular}{l l c c c c c} 
\hline
\hline
 & & \multicolumn{5}{c}{Dependent variable: Likelihood of Vote for 'Test'} \\ \cmidrule(l){3-7} 
 & & (1) & (2) & (3) & (4) & (5) \\ 
\midrule
 \multicolumn{2}{l}{Own Performance Prediction} & 0.054$^{***}$ & 0.049$^{***}$ & 0.054$^{***}$ & 0.061$^{***}$ & 0.051$^{***}$ \\ 
  & & (0.000) & (0.000) & (0.000) & (0.000) &  (0.000) \\ 
 \multicolumn{2}{l}{Group Performance Prediction} & -0.052$^{***}$ & -0.047$^{***}$ & -0.051$^{***}$ &  -0.053$^{***}$ & -0.046$^{***}$ \\ 
  & & (0.000) & (0.000) & (0.000) & (0.000) & (0.000) \\ 
  \multicolumn{2}{l}{Sample Performance} & 0.105$^{***}$ & 0.127$^{***}$ &  0.113$^{***}$ & 0.089$^{**}$ &  0.126$^{***}$ \\ 
  & & (0.000) & (0.000) & (0.000) & (0.003) &  (0.000) \\
  & & &  & &  &  \\
  \multicolumn{2}{l}{Age} & & 0.002 &  &  & 0.002\\ 
  & & & (0.460) &  &  & (0.490) \\ 
 \multicolumn{2}{l}{Female} &  & -0.167$^{***}$ &  &   & -0.180$^{***}$ \\ 
  & & & (0.000) &  &   & (0.000) \\ 
 \multicolumn{2}{l}{Student} &  & 0.126$^{*}$ &   &  & 0.155$^{**}$\\ 
  & & & (0.020) &  &  & (0.003) \\ 
 \multicolumn{2}{l}{Degree in Higher Education} &  & 0.022 &  &   & 0.066 \\ 
  & & & (0.620) &  &  & (0.146) \\ 
  & & &  & &  &  \\
  \multicolumn{2}{l}{Sure to have made the right Vote} &  &  &  -0.017 &  & -0.007 \\ 
  & & &  & (0.323) &  & (0.671) \\ 
  \multicolumn{2}{l}{Difficulty to change Vote} &  &  & 0.006 &  & 0.002  \\ 
  & & &  & (0.698) &  &  (0.866) \\ 
  \multicolumn{2}{l}{Importance of doing well in Test} &  &  & 0.015 &  &  0.028 \\ 
  & & &  & (0.358) &  &  (0.122) \\ 
  \multicolumn{2}{l}{Expected Test Difficulty} &  &  & 0.040 &  & 0.038 \\ 
  & & &  & (0.090) &  &  (0.112) \\ 
  \multicolumn{2}{l}{Expectation of doing well in Test} &  &  & 0.016 &  &  0.030 \\ 
  & & &  & (0.507) &  &  (0.169) \\ 
  & & &  & &  &  \\
  \multicolumn{2}{l}{Social Comparison Orientation (INCOM)} &  &  &   & -0.069$^{*}$ & -0.064$^{*}$ \\ 
  & & &  &   & (0.021) & (0.029) \\ 
 \multicolumn{2}{l}{General Self-Efficacy} &  &  &  & -0.053 & -0.068\\ 
  & & &  &  & (0.110) & (0.059) \\ 
 \multicolumn{2}{l}{Altruism} &  &  &   & -0.022 & -0.018 \\ 
  & & &  &  & (0.548) & (0.652) \\ 
 \multicolumn{2}{l}{Risk Attitude} &  &  &  & -0.009 & -0.013 \\ 
  & & &  &  & (0.364) & (0.183) \\   
 \multicolumn{2}{l}{Ambiguity Attitude} &  &  &  & 0.001 & 0.002 \\ 
  & & &  &  &  (0.757) & (0.528) \\ 
\hline
\hline
\end{tabular} 
\begin{tablenotes}
    \small
    \item \textit{Note:} p-values in parentheses; $^{*}\, p<0.05$; $^{**}\, p<0.01$; $^{***}\, p<0.001$.
\end{tablenotes}
\end{threeparttable} 
\end{table}

\clearpage

\begin{table}[H]
    \caption{Aspect by aspect comparison of experimental designs}
    \label{tab:comparison}
    \centering
    \begin{tabular}{|p{7.5cm}|p{7.5cm}|}
        \hline
        \multicolumn{1}{|>{\centering\arraybackslash}m{75mm}|}{\textbf{Hoelzl \& Rustichini (2005)}} & \multicolumn{1}{>{\centering\arraybackslash}m{75mm}|}{\textbf{Protte (2025)}} \\
         \hline        
         \textbf{Environment} &  \\ [.5ex]
         \textbullet\ Laboratory & \textbullet\ Remote \\ [.5ex]
         \textbullet\ Paper \& pen & \textbullet\ Computerized \\ [.5ex]
         & \\
         \textbf{Recruitment} &  \\ [.5ex]
         \textbullet\ University campus & \textbullet\ Prolific.com \\ [1ex]
         & \\
         \textbf{Sample} &  \\ [.5ex]
         \textbullet\ N = 134 & \textbullet\ N = 200 (182 in analysis)  \\ [.5ex]
         \textbullet\ Mostly university students (plus 6 professionals, 4 pupils) & \textbullet\ Heterogeneous sample (32\% students) \\ [.5ex]
         \textbullet\ ø age: 23 & \textbullet\ ø age: 34.3 \\ [1ex]
         & \\
         \textbf{Treatments} & \\ [.5ex]
         \textbullet\ Incentive (money vs. no money) & \textbullet\ Lottery mechanism (probabilistic outcome distribution vs. fixed outcome distribution) \\ [.5ex]
         \textbullet\ Task difficulty (hard vs. easy) &  \\ [.5ex]
         & \\
         \textbf{Session sizes} &  \\ [.5ex]
         \textbullet\ Multiple groups of "at least 9" subjects, one group of 7 & \textbullet\ Two groups of 100 subjects\\ [.5ex]
         & \\
         \textbf{Indicator variable} &  \\ [.5ex]
         \textbullet\ Vote on payoff mechanism & \textbullet\ Vote on payoff mechanism \\ [.5ex]
         & \\
         \textbf{Task} &  \\ [.5ex]
         \textbullet\ LEWITE (20 items) & \textbullet\ SAT Analogies (20 items)  \\ [.5ex]
         \textbullet\ Choose 2 from 7-9 alternatives & \textbullet\ Choose 1 from 5 alternatives \\ [.5ex]
         & \\
         \textbf{Payment} & \\ [.5ex]
         \textbullet\ Fixed payment: None & Fixed payment: 3.00 EUR ($\sim3.30$ USD) \\ [.5ex]
         \textbullet\ Bonus: 150 ATS (\~10 USD at the time) & Bonus: 5.00 EUR ($\sim5.50$ USD) \\ [.5ex]
         & \\
         \textbf{Procedure} &  \\ [.5ex]
         \textbullet\ Instructions distributed and read out aloud & \textbullet\ Instructions presented on screen \\ [.5ex]
          & \textbullet\ Comprehension questions \\ [.5ex]
         \textbullet\ Practice examples and solutions & \textbullet\ Practice examples and solutions \\ [.5ex]
         \textbullet\ Vote & \textbullet\ Vote \& Voting rationale \\ [.5ex]
         \textbullet\ Pre-task questions and predictions & \textbullet\ Pre-task questions and predictions \\ [.5ex] 
         \textbullet\ Test (results revealed to subjects but otherwise kept confidential)  & \textbullet\ Test (results revealed to subjects but otherwise kept confidential) \\ [.5ex]
         \textbullet\ Individual die roll & \textbullet\ Post-task questions and estimations \\ [.5ex] 
         \textbullet\ Post-task questions and estimations & \textbullet\ Choice of lucky number (die roll equivalent) \\ [.5ex]
          & \textbullet\ Additional questionnaires (INCOM, GSE, HEXACO altruism, MPL risk, MPL ambiguity, demographics) \\ [.5ex]         
         \textbullet\ Voting result \& median test result announced verbally & \textbullet\ Voting result \& median test result announced via bulk mail \\ [.5ex]  
        \hline
    \end{tabular}
\end{table}

\clearpage
\section{Experimental design} \label{instructions}

\textit{[Translated from German]}

\section*{Welcome!}

Thank you for participating in today’s study!

This is a non-commercial study conducted by researchers at Paderborn University and funded by Paderborn University.

Please note the following information:

\begin{itemize}
    \item The data collected through this study is completely anonymous and is used exclusively for scientific purposes. It is not possible to draw conclusions about you or other persons.
    \item Your data will be conscientiously protected in accordance with the provisions of the European General Data Protection Regulation (GDPR), even in the case of a scientific publication.
    \item Your participation in the study is voluntary and can be terminated at any time without giving reasons.
\end{itemize}

Please answer the questions as carefully and conscientiously as possible.
Please only start working on the study when you have enough time to answer them in one sitting. Communication programs (e.g. chat or e-mail) should be closed during processing in order to avoid distractions.

Please press “Start experiment” to acknowledge that you have read and agreed with the conditions as
stated above. If you do not agree with the conditions, please close your browser window.

\begin{center}
***
\end{center}

\section*{Instructions}

Please read the following instructions carefully, as you will be asked comprehension questions on the next page!\\

You will receive a fixed payment of £2.00 for your participation.
In addition, you can receive a bonus payment of £5.00. The conditions for this bonus payment are explained below.\\

Not all participants will receive a bonus payment.
Whether you receive a bonus payment is decided either by (a) a performance test or (b) a lottery.

In order to decide which payoff mechanism (performance test or lottery) determines whether you receive a bonus payment, all participants in this experiment will vote on the payoff mechanism at the beginning:

Under the "performance test" condition, you will receive a bonus payment if your score is in the upper half of all 100 participants, i.e. if your performance is among the top 50 participants.

[Replication group] Under the "Lottery" condition, you choose a lucky number between 1 and 6. After all 100 people have taken part, a random generator throws a conventional 6-sided dice. If the dice shows an odd number (i.e. 1, 3, 5), you will receive a bonus payment if you have entered an odd number as your lucky number. If the dice shows an even number (i.e. 2, 4, 6), you will receive a bonus payment if you have entered an even number as your lucky number.

[Adaptation group] Under the "Lottery" condition, you choose a lucky number between 1 and 6, which, together with your Prolific ID, forms your winning code. After all 100 people have taken part, a random generator draws 50 winning codes. If your winning code is drawn, you will receive a bonus payment.\\

In total, 100 people will participated in this study. A simple majority will decide by secret vote which payoff mechanism will be applied.

If the majority vote in favor of the "performance test" payoff mechanism, your test result will determine whether you receive a bonus payment. If the majority of participants vote in favor of the "lottery" payoff mechanism, the lottery result will decide whether you receive a bonus payment.
(If there is an exact tie, a coin toss by the experimenters will decide).

As you do not yet know at the time of your participation which payout mechanism has ultimately been voted, you will play both the test and the lottery. Only you will be informed of your individual test result.

As soon as all 100 people have taken part in this study, you will be informed via an email to your Prolific ID about which payment mechanism has been voted and whether you will receive a bonus payment. You will also be informed about the average test performance of all participants.\\

Please make sure that you have read and understood the instructions, as you will be asked some comprehension questions on the following page. Following the comprehension questions, you will receive three examples of the test tasks.

After the experiment section, we will ask you to complete questionnaire in which there are no "right" or "wrong" answers. Please simply answer the questions according to your personal judgments and opinions.\\

\textit{(The "Next" button appears after 90 seconds.)}

\begin{center}
***
\end{center}

\section*{Comprehension questions}

\textit{[Question 5 differed between treatments, 'X' indicates correct answer]}

1) Which of the following statements about the procedure of the experiment is correct?
\begin{table}[H]
    \begin{tabular}{c l}
         $\ocircle$ & You will play both the test and the lottery, regardless of which you choose in the vote. X\\
         $\ocircle$ & You will either only play the test or only the lottery and decide in the vote.\\
    \end{tabular}
    \label{tab:comp1}
\end{table}

2) Which of the following statements about the payoff mechanism is correct?
\begin{table}[H]
    \begin{tabular}{c l}
         $\ocircle$ & You alone decide which payout mechanism is used for you. \\
         $\ocircle$ & A majority vote among all participants determines which payout mechanism will be used for all \\
         & participants. X\\
    \end{tabular}
    \label{tab:comp2}
\end{table}

3) Which of the following statements about the bonus payment is correct?
\begin{table}[H]
    \begin{tabular}{c l}
         $\ocircle$ & All participants receive a bonus payment. \\
         $\ocircle$ & The test result or the lottery result (depending on the voting result) decides who receives a bonus \\
         & payment. X \\
    \end{tabular}
    \label{tab:comp3}
\end{table}

4) Which of the following statements about the performance test is correct?
\begin{table}[H]
    \begin{tabular}{c l}
         $\ocircle$ & If the performance test is voted as the payoff mechanism, 50 randomly selected participants will \\ 
         & receive a bonus payment. \\
         $\ocircle$ & If the performance test is voted as the payoff mechanism, half of the participants (the 50 with the \\
         & best test results) will receive a bonus payment. X\\
    \end{tabular}
    \label{tab:comp4}
\end{table}

5 - Replication group) Which of the following statements about the lottery is correct?
\begin{table}[H]
    \begin{tabular}{c l}
         $\ocircle$ & If the lottery is voted as the payoff mechanism, you have a 50\% probability of receiving a bonus \\
         & payment (namely if you have selected an odd lucky number and an odd number is rolled or if you \\
         & have selected an even lucky number and an even number is rolled). X\\
         $\ocircle$ & If the lottery is voted as the payout mechanism, the probability of not receiving a bonus payment \\
         & is higher than the probability of receiving a bonus payment.\\
    \end{tabular}
    \label{tab:comp5rep}
\end{table}

5 - Adaptation group) Which of the following statements about the lottery is correct?
\begin{table}[H]
    \begin{tabular}{c l}
         $\ocircle$ & If the lottery is voted as the payoff mechanism, exactly half of the participants (50 out of 100) \\
         & will receive a bonus payment (i.e. if their winning code is drawn). X \\
         $\ocircle$ & If the lottery is voted as the payoff mechanism, more participants will receive a bonus payment \\
         & than in the performance test. \\
    \end{tabular}
    \label{tab:comp5adapt}
\end{table}

6) When will you be informed about your test result?
\begin{table}[H]
    \begin{tabular}{c l}
         $\ocircle$ & Immediately after completing the test. X \\
         $\ocircle$ & After all 100 people participated in the study. \\
    \end{tabular}
    \label{tab:comp6}
\end{table}

7) When will you be informed whether you will receive a bonus payment?
\begin{table}[H]
    \begin{tabular}{c l}
         $\ocircle$ & Immediately after completing the test. \\
         $\ocircle$ & After all 100 people participated in the study. X \\
    \end{tabular}
    \label{tab:comp7}
\end{table}

\begin{center}
***
\end{center}

\clearpage
\section*{Vote} \label{vote}

Your task in the test is to identify the most appropriate analogy between pairs of words. Below you can see three examples of test tasks.

In total, the test consists of \textbf{20 tasks}.

Please select one of the five possible answers each.
The correct solution will be displayed on the next page.

Which pair of words relates to each other like:

\textbf{1) "seed" to "plant"}

$\ocircle$ pouch - kangaroo $\ocircle$ root - soil $\ocircle$ drop - water $\ocircle$ bark - tree $\ocircle$ \underline{egg - bird}

\textbf{2) "unfetter" to "pinioned"}

$\ocircle$ recite - practiced 
$\ocircle$ sully - impure  
$\ocircle$ \underline{enlighten - ignorant } 
$\ocircle$ revere - unrecognized 
$\ocircle$ adore - cordial

\textbf{3) "vacillate" to "indecision"}

$\ocircle$ \underline{lament - woe} $\ocircle$ hibernate - winter $\ocircle$ extricate - entanglements $\ocircle$ digress - angst $\ocircle$ emulate - egotism \\

\textit{[Correct solutions underlined here for conciseness.]}

\begin{center}
***
\end{center}

Now, please cast your vote for the choice of payoff mechanism and then click "Confirm".\\

\textbf{Your vote:}

$\ocircle$ Test
$\ocircle$ Lottery \\

\textit{(The "Next" button appears after 30 seconds.)}

\begin{center}
***
\end{center}

Please explain the rationale for your choice in 1-2 sentences: \_\_\_\_\_

\begin{center}
***
\end{center}

\clearpage
\section*{Performance test} \label{test}

Please indicate your expectations for the following aspects:\\

\textbf{How sure are you that you have made the right decision in the vote?}

Very unsure $\ocircle$ $\ocircle$ $\ocircle$ $\ocircle$ $\ocircle$ $\ocircle$ $\ocircle$ Very sure

\textbf{How difficult would you find it to change your decision?}

Very easy $\ocircle$ $\ocircle$ $\ocircle$ $\ocircle$ $\ocircle$ $\ocircle$ $\ocircle$ Very difficult

\textbf{How important is doing well in the test for you?}

Not important at all $\ocircle$ $\ocircle$ $\ocircle$ $\ocircle$ $\ocircle$ $\ocircle$ $\ocircle$ Very important 

\textbf{How difficult do you expect the test to be?}

Very easy $\ocircle$ $\ocircle$ $\ocircle$ $\ocircle$ $\ocircle$ $\ocircle$ $\ocircle$ Very difficult 

\textbf{How do you think you will do in the test?}

Very bad $\ocircle$ $\ocircle$ $\ocircle$ $\ocircle$ $\ocircle$ $\ocircle$ $\ocircle$ Very good \\

Please specify:\\

How many of the 20 questions will \textbf{you} answer correctly? \\

How many of the 20 questions will \textbf{the other participants} answer correctly \textbf{on average}? \\

\begin{center}
***
\end{center}

Thank you for your participation on this experiment so far!

The announced test will begin on the following page.

Please only advance when you have enough time to complete the tasks.

The test consists of 20 tasks in total. You have 30 seconds per task. 

At the top of the screen you will see a timer that starts automatically as soon as you begin the test.

\begin{center}
***
\end{center}

Please select one of the five possible answers and then click "Next".

\begin{center}
    [Items can be obtained upon request.]
\end{center}

\begin{center}
***
\end{center}

Please specify:\\

How many of the 20 questions did \textbf{the other participants} answer correctly \textbf{on average}? \\

In retrospect, please assess the following aspects:\\

\textbf{How satisfied are you with your test result?}

Not satisfied at all $\ocircle$ $\ocircle$ $\ocircle$ $\ocircle$ $\ocircle$ $\ocircle$ $\ocircle$ Very satisfied

\textbf{How sure are you now that you have made the right decision in the vote?}

Very unsure $\ocircle$ $\ocircle$ $\ocircle$ $\ocircle$ $\ocircle$ $\ocircle$ $\ocircle$ Very sure

\textbf{How difficult would you find it now to change your decision?}

Very easy $\ocircle$ $\ocircle$ $\ocircle$ $\ocircle$ $\ocircle$ $\ocircle$ $\ocircle$ Very difficult

\textbf{How difficult did you find the test?}

Very easy $\ocircle$ $\ocircle$ $\ocircle$ $\ocircle$ $\ocircle$ $\ocircle$ $\ocircle$ Very difficult

\begin{center}
***
\end{center}

\clearpage
\section*{Lottery} \label{lottery}

You have now completed the test. Thank you for participating in the experiment so far!

On the following page we will continue with the lottery.

\begin{center}
***
\end{center}

\textit{[Lottery Replication]}

Reminder:\\
In this lottery, you choose a lucky number between 1 and 6.
After all 100 people have taken part, a random number generator throws a conventional \textbf{6-sided dice}.\\

If the dice shows an odd number (i.e. 1, 3, 5), you will receive a bonus payment if you have entered an odd number as your lucky number.\\
If the dice shows an even number (i.e. 2, 4, 6), you will receive a bonus payment if you have entered an even number as your lucky number.\\

Please enter your lucky number here: \_\_\_ \\

\textit{[Lottery Adaption]}

Reminder:\\
In this lottery, you choose a lucky number between 1 and 6. This, together with your Prolific-ID, forms your \textbf{winning code}.\\

After all 100 people have taken part, a random generator draws 50 winning codes. If your winning code is drawn, you will receive a bonus payment.\\

Please enter your lucky number here: \_\_\_ \\

\begin{center}
***
\end{center}

\clearpage
\section*{Questionnaire} \label{questionnaire}

Thank you for completing the experiment so far!

In the following, we ask you to complete a multi-part questionnaire. There are no "right" or "wrong" answers.
Simply indicate to what extent the following statements apply to you personally.

\begin{center}
***
\end{center}

\begin{table}[H]
    \centering
    \begin{adjustbox}{max width=\textwidth}
    \begin{tabular}{l c c c c c}
    \hline
        & \makecell{Strongly \\ disagree} & \makecell{Disagree}  & \makecell{Neither agree\\ nor disagree } & \makecell{Agree} & \makecell{Strongly \\ agree} \\
    \hline
         I always pay a lot of attention to how I do things compared with how others do things. & $\ocircle$ & $\ocircle$ & $\ocircle$ & $\ocircle$ & $\ocircle$ \\
         I often compare how I am doing socially (e.g., social skills, popularity) with other people. & $\ocircle$ & $\ocircle$ & $\ocircle$ & $\ocircle$ & $\ocircle$ \\
         I am not the type of person who compares often with others. & $\ocircle$ & $\ocircle$ & $\ocircle$ & $\ocircle$ & $\ocircle$ \\
         Please choose “Strongly disagree”. & $\ocircle$ & $\ocircle$ & $\ocircle$ & $\ocircle$ & $\ocircle$ \\
         I often try to find out what others think who face similar problems as I face. & $\ocircle$ & $\ocircle$ & $\ocircle$ & $\ocircle$ & $\ocircle$ \\
         I always like to know what others in a similar situation would do. & $\ocircle$ & $\ocircle$ & $\ocircle$ & $\ocircle$ & $\ocircle$ \\
         If I want to learn more about something, I try to find out what others think about it. & $\ocircle$ & $\ocircle$ & $\ocircle$ & $\ocircle$ & $\ocircle$ \\
    \hline
    \end{tabular}
    \end{adjustbox}
    \label{tab:incom}
\end{table}

\begin{center}
***
\end{center}

\begin{table}[H]
    \centering
    \begin{adjustbox}{max width=\textwidth}    
    \begin{tabular}{l c c c c c}
    \hline
        &  \makecell{Does not apply\\ at all} & \makecell{Applies \\little} & \makecell{Somewhat \\ applies} & \makecell{Pretty much \\ applies} & \makecell{Fully \\ applies} \\
    \hline
         In difficult situations I can rely on my abilities. & $\ocircle$ & $\ocircle$ & $\ocircle$ & $\ocircle$ & $\ocircle$ \\
         I can cope well with most of the problems on my own power. & $\ocircle$ & $\ocircle$ & $\ocircle$ & $\ocircle$ & $\ocircle$ \\
         Even strenuous and complicated tasks I can usually solve well. & $\ocircle$ & $\ocircle$ & $\ocircle$ & $\ocircle$ & $\ocircle$ \\
    \hline
    \end{tabular}
    \end{adjustbox}    
    \label{tab:asku}
\end{table}

\begin{center}
***
\end{center}

\begin{table}[H]
    \centering
    \begin{adjustbox}{max width=\textwidth}
    \begin{tabular}{l c c c c c}
    \hline
        & \makecell{Strongly \\ agree} & \makecell{Agree}  & \makecell{Neither degree \\ nor disagree} & \makecell{Disagree} & \makecell{Strongly \\ disagree} \\
    \hline
        I have sympathy for people who are less fortunate than I am. & $\ocircle$ & $\ocircle$ & $\ocircle$ & $\ocircle$ & $\ocircle$\\
        I try to give generously to those in need.  & $\ocircle$ & $\ocircle$ & $\ocircle$ & $\ocircle$ & $\ocircle$\\
        It wouldn’t bother me to harm someone I didn’t like. & $\ocircle$ & $\ocircle$ & $\ocircle$ & $\ocircle$ & $\ocircle$\\
        People see me as a hard-hearted person. & $\ocircle$ & $\ocircle$ & $\ocircle$ & $\ocircle$ & $\ocircle$\\ 
    \hline
    \end{tabular}
    \end{adjustbox}
    \label{tab:altruism}
\end{table}

\begin{center}
***
\end{center}

\clearpage
\subsection*{Please see the following instructions.}

The following table contains 29 separate decisions.\\

Please decide for each row which lottery you prefer: "Option A" or "Option B". \\

To specify your preferences between Option A and Option B, you only need to make one choice, namely in which row you would like to switch \textbf{from  Option A to Option B}.\\

For each row \textbf{before} the one you have selected, you prefer Option A.
For each row \textbf{after} the one you have selected, including the one you have selected, you prefer Option B.

\begin{table}[H]
    \centering
    \begin{tabular}{| l | c | c |}
    \hline
        & \textbf{Option A} & \textbf{Option B} \\
    \hline
        1) & €30 with a probability of 50\%, €0 with a probability of 50\% & €1 safe\\
        2) & €30 with a probability of 50\%, €0 with a probability of 50\% & €2 safe \\
        3) & €30 with a probability of 50\%, €0 with a probability of 50\% & €3 safe \\
        4) & €30 with a probability of 50\%, €0 with a probability of 50\% & €4 safe \\
        5) & €30 with a probability of 50\%, €0 with a probability of 50\% & €5 safe \\
        6) & €30 with a probability of 50\%, €0 with a probability of 50\% & €6 safe \\
        7) & €30 with a probability of 50\%, €0 with a probability of 50\% & €7 safe \\
        8) & €30 with a probability of 50\%, €0 with a probability of 50\% & €8 safe \\
        9) & €30 with a probability of 50\%, €0 with a probability of 50\% & €9 safe \\
        10) & €30 with a probability of 50\%, €0 with a probability of 50\% & €10 safe \\
        11) & €30 with a probability of 50\%, €0 with a probability of 50\% & €11 safe \\
        12) & €30 with a probability of 50\%, €0 with a probability of 50\% & €12 safe \\
        13) & €30 with a probability of 50\%, €0 with a probability of 50\% & €13 safe \\
        14) & €30 with a probability of 50\%, €0 with a probability of 50\% & €14 safe \\
        15) & €30 with a probability of 50\%, €0 with a probability of 50\% & €15 safe \\
        16) & €30 with a probability of 50\%, €0 with a probability of 50\% & €16 safe \\
        17) & €30 with a probability of 50\%, €0 with a probability of 50\% & €17 safe \\
        18) & €30 with a probability of 50\%, €0 with a probability of 50\% & €18 safe \\
        19) & €30 with a probability of 50\%, €0 with a probability of 50\% & €19 safe \\
        20) & €30 with a probability of 50\%, €0 with a probability of 50\% & €20 safe \\
        21) & €30 with a probability of 50\%, €0 with a probability of 50\% & €21 safe \\
        22) & €30 with a probability of 50\%, €0 with a probability of 50\% & €22 safe \\
        23) & €30 with a probability of 50\%, €0 with a probability of 50\% & €23 safe \\
        24) & €30 with a probability of 50\%, €0 with a probability of 50\% & €24 safe \\
        25) & €30 with a probability of 50\%, €0 with a probability of 50\% & €25 safe \\
        26) & €30 with a probability of 50\%, €0 with a probability of 50\% & €26 safe \\
        27) & €30 with a probability of 50\%, €0 with a probability of 50\% & €27 safe \\
        28) & €30 with a probability of 50\%, €0 with a probability of 50\% & €28 safe \\
        29) & €30 with a probability of 50\%, €0 with a probability of 50\% & €29 safe \\
    \hline
    \end{tabular}
    \label{tab:mpl}
\end{table}

Please indicate the row for which you want to switch from Option A to Option B: \_\_\_

\textit{(The "Next" button appears after 60 seconds.)}

\begin{center}
***
\end{center}

\clearpage
\subsection*{Please see the following instructions.} 

The following table contains 20 separate decisions.\\
First, please choose your success color: $\ocircle$ \textbf{Red}   $\ocircle$ \textbf{Black} \\

Afterwards, please decide for each row which lottery you prefer: "Urn A" or "Urn B". 
Both urns contain exactly 100 balls.\\
Urn A contains the exact same number of Red and Black balls (50 each).\\
The distribution of Red and Black balls in Urn B is unknown.\\

To specify your preferences between Urn A and Urn B, you only need to make one choice, namely in which row you would like to switch \textbf{from Urn A to Urn B}.

For each row \textbf{before} the one you have selected, a ball is drawn from Urn A.
For each row \textbf{after} the one you have selected, including the one you have selected, Urn B is used for the draw.

Exactly one ball will be drawn. You will receive...

\begin{table}[H]
    \centering
    \begin{tabular}{| l | c | c |}
    \hline
        & \textbf{Urn A} & \textbf{Urn B} \\
        & \textbf{50 Red balls, 50 Black balls} & \textbf{? Red balls, ? Black balls} \\        
    \hline
        1) & €20.00 if Chosen Color & €16.40 if Chosen Color \\
         & €0 if not & €0 if not \\
        2) & €20.00 if Chosen Color & €17.20 if Chosen Color \\
         & €0 if not & €0 if not \\
        3) & €20.00 if Chosen Color & €18.00 if Chosen Color \\
         & €0 if not & €0 if not \\
        4) & €20.00 if Chosen Color & €18.80 if Chosen Color \\
         & €0 if not & €0 if not \\
        5) & €20.00 if Chosen Color & €19.60 if Chosen Color \\
         & €0 if not & €0 if not \\
        6) & €20.00 if Chosen Color & €20.40 if Chosen Color \\
         & €0 if not & €0 if not \\
        7) & €20.00 if Chosen Color & €21.20 if Chosen Color \\
         & €0 if not & €0 if not \\
        8) & €20.00 if Chosen Color & €22.00 if Chosen Color \\
         & €0 if not & €0 if not \\
        9) & €20.00 if Chosen Color & €22.80 if Chosen Color \\
         & €0 if not & €0 if not \\
        10) & €20.00 if Chosen Color & €23.60 if Chosen Color \\
         & €0 if not & €0 if not \\
        11) & €20.00 if Chosen Color & €24.40 if Chosen Color \\
         & €0 if not & €0 if not \\
        12) & €20.00 if Chosen Color & €25.20 if Chosen Color \\
         & €0 if not & €0 if not \\
        13) & €20.00 if Chosen Color & €26.00 if Chosen Color \\
         & €0 if not & €0 if not \\
        14) & €20.00 if Chosen Color & €26.80 if Chosen Color \\
         & €0 if not & €0 if not \\
        15) & €20.00 if Chosen Color & €27.60 if Chosen Color \\
        & €0 if not & €0 if not \\
        16) & €20.00 if Chosen Color & €28.40 if Chosen Color \\
         & €0 if not & €0 if not \\
        17) & €20.00 if Chosen Color & €29.20 if Chosen Color \\
         & €0 if not & €0 if not \\
        18) & €20.00 if Chosen Color & €30.00 if Chosen Color \\
         & €0 if not & €0 if not \\
        19) & €20.00 if Chosen Color & €30.80 if Chosen Color \\
         & €0 if not & €0 if not \\
        20) & €20.00 if Chosen Color & €31.60 if Chosen Color \\
         & €0 if not & €0 if not \\
    \hline
    \end{tabular}
    \label{tab:aa}
\end{table}

Please indicate the decision for which you want to switch from Urn A to Urn B: \_\_\_

\textit{(The "Next" button appears after 60 seconds.)}

\begin{center}
***
\end{center}

\clearpage
\subsection*{Please answer the following questions.} \label{quest_demographics}

What is your age? \\

What is your gender?
\begin{table}[H]
    \begin{tabular}{c l}
         $\ocircle$ & Male \\
         $\ocircle$ & Female \\
         $\ocircle$ & Non-binary \\
    \end{tabular}
    \label{tab:gender}
\end{table}

Please indicate the highest educational qualification you have obtained to date:
\begin{table}[H]
    \begin{tabular}{c l}
         $\ocircle$ & Middle school \\
         $\ocircle$ & High school/A-levels \\
         $\ocircle$ & Vocational training \\
         $\ocircle$ & Undergraduate degree \\
         $\ocircle$ & Graduate degree \\         
         $\ocircle$ & Ph.D. or higher \\                  
    \end{tabular}
    \label{tab:major}
\end{table}

Are you currently a university student?
\begin{table}[H]
    \begin{tabular}{c l}
         $\ocircle$ & Yes \\
         $\ocircle$ & No \\
    \end{tabular}
    \label{tab:student}
\end{table}

What is your current study major or profession? \\

Is German your first language?
\begin{table}[H]
    \begin{tabular}{c l}
         $\ocircle$ & Yes \\
         $\ocircle$ & No \\
    \end{tabular}
    \label{tab:language}
\end{table}

In general, how willing are you to take risks?

Not at all willing to take risks $\ocircle$ $\ocircle$ $\ocircle$ $\ocircle$ $\ocircle$ $\ocircle$ $\ocircle$ $\ocircle$ $\ocircle$ $\ocircle$ $\ocircle$ Very willing to take risks \\

Is there anything else you would like to tell us? (optional)

\begin{center}
***
\end{center}

\clearpage
\section*{Thank you for your participation!} \label{final_pages}

Almost done...

For the evaluation of the experiment, it is important that we can rely on you having completed the experiment attentively and honestly.

If you did not complete the experiment attentively and honestly or if you completed it with interruptions, please let us know in the following questions.

Don't worry, you will still receive your reward!

Did you work through the experiment carefully?
\begin{table}[H]
    \begin{tabular}{c l}
         $\ocircle$ & Yes, I worked on the experiment carefully. \\
         $\ocircle$ & No, I did NOT work on the experiment carefully. \\
    \end{tabular}
    \label{tab:attention}
\end{table}

Did you complete the experiment in one go, i.e. without interruptions?
\begin{table}[H]
    \begin{tabular}{c l}
         $\ocircle$ & Yes, I completed the experiment in one go. \\
         $\ocircle$ & No, I did NOT complete  the experiment in one go. \\
    \end{tabular}
    \label{tab:attention2}
\end{table}

\begin{center}
***
\end{center}

\textbf{You have completed the experiment!}

We will announce the winners of the bonus payment, as well as which payment mechanism was chosen for this, by circular mail to your Prolific accounts as soon as the number of participants reaches 100.

We will also inform you of your ranking in the test (only your Prolific ID will be displayed).

Thank you for your participation!

\begin{center}
***
\end{center}

\end{document}